\documentclass[12pt]{article}
\usepackage{graphicx,epsfig,here}    
\begin{document}

\renewcommand{\refname}{}
\newcommand{\vsp}{\vspace*{-0.3cm}}

\begin{center}
\thispagestyle{empty}
\vspace*{3cm}

{\Huge \bf HadAtom02}

\vspace{7mm}

Workshop on Hadronic Atoms\\
CERN, CH-1211 Geneva 23, Switzerland\\
October 14 -- 15, 2002\\
\vspace*{12mm}

L.~Afanasyev\footnote{E-mail: Leonid.Afanasev@cern.ch}\\ 
$^1$ {\em Joint Institute for Nuclear Research, Dubna, Moscow Region, 141980 Russia}

\vspace{5mm}
A.~Lanaro\footnote{E-mail: Armando.Lanaro@cern.ch}\\
{\em CERN, CH-1211 Geneva 23, Switzerland}

\vspace{5mm}
J.~Schacher\footnote{E-mail: schacher@lhep.unibe.ch}\\
{\em Lab. for High-Energy Physics, University of Bern,
Sidlerstrasse 5, CH-3012, Bern, Switzerland}
\end{center}

\vspace{10mm}
\begin{abstract}
  \baselineskip 1.5em These are the proceedings of the workshop
  "HadAtom02", held at the CERN, October 14 - 15, 2002. The main topic
  of the workshop concerned the physics of hadronic atoms and in this
  contest recent results from experiments and theory were presented.
  These proceedings contain the list of participants, the scientific
  program and a short contribution from each speaker.
\end{abstract}

\newpage
\section{Introduction}
The workshop ``HadAtom02'' took place at CERN on October 14--15, 2002.
It was the 4th in a series of workshops on bound states, in particular
hadronic atoms. The previous ones were held in Dubna, Russia (May
1998) [1] and in Bern (October 1999, October 2001) [2,3].  The meeting
was attended by about 40 physicists, and contributions were presented
by 17 participants.

The workshop programme included:

\vspace*{2mm}

\noindent
\begin{tabular}{l@{\hspace{5mm}}l}
$\bullet$ Hadronic atoms, in particular their & $\bullet$ Experiments \\ 
\hspace*{4mm} Production & \hspace*{4mm} DIRAC at CERN \\ 
\hspace*{4mm} Interaction with matter & \hspace*{4mm} DEAR at DAFNE \\ 
\hspace*{4mm} Energy levels & 
\hspace*{4mm} PSI (Pionic Hydrogen Collaboration) \\
\hspace*{4mm} Decays & \hspace*{4mm} NA48 \\ 
$\bullet$ Meson-meson and meson-baryon scattering & \hspace*{4mm} Others \\ 
&
$\bullet$ $K_{\ell 4}$ decays \\
\end{tabular}

\vspace*{2mm}

The talks were devoted to recent experimental and theoretical progresses
in the investigations of hadronic atoms. Among the highlights of the
workshop were the presentation of pre\-limi\-nary results from the DIRAC
collaboration on the measurement of the lifetime of pionium, as well
as the first measurement of kaonic nitrogen by the DEAR collaboration.

The speakers have provided a two-page summary of the presentations
including a list of the most relevant references --- these
contributions are collected below.  These proceedings include also a list of the
participants with their e-mail, and the scientific program of the
workshop.

\vskip.5cm

{\bf Acknowledgment}

We would like to thank all participants for their effort to travel to
CERN and for making ``HadAtom02'' an exciting and lively meeting.  We
furthermore thank our secretary Catherine Moine for her professional
contribution to the organization of the meeting.  Last but not least,
we thank our colleagues from the organizing committee, Juerg Gasser,
Leonid Nemenov, Akaki Rusetsky, and Dirk Trautmann for their
invaluable contribution in structuring the meeting.

\vspace*{.7cm}

 CERN, January 2002

\vskip.5cm

Leonid Afanasyev, Armando Lanaro and J\"urg Schacher

\bigskip

\noindent\hrulefill

\begin{itemize}

\item[{[1]}]
Proceedings of the International Workshop ``Hadronic Atoms and Positronium in
the Standard Model'', Dubna, 26-31 May 1998, ed. M.A. Ivanov, A.B. Arbuzov,
E.A. Kuraev, V.E. Lyubovitskij, A.G. Rusetsky.
 
\vsp
\item[{[2]}]
J.~Gasser, A.~Rusetsky and J.~Schacher,
arXiv:hep-ph/9911339.

\vsp
\item[{[3]}]
J.~Gasser, A.~Rusetsky and J.~Schacher,
arXiv:hep-ph/0112293.

\end{itemize}

\newpage
\section{List of participants}
\begin{tabular}{rll}
 1. & Afanasyev Leonid (Dubna) & Leonid.Afanasev@cern.ch\\
 2. & B\"uttiker Paul (Orsay) & buttiker@ipno.in2p3.fr\\
 3. & Baur Gerhard (Juelich) & g.baur@fz-juelich.de\\
 4. & Cheshkov Cvetan (Saclay) & cvetan@hep.saclay.cea.fr\\
 5. & Colangelo Gilberto (Bern) & gilberto@itp.unibe.ch\\
 6. & Descotes-Genon Sebastien (Southampton) & sdg@hep.phys.soton.ac.uk\\
 7. & Ericson Magda (CERN) & Magda.Ericson@cern.ch\\
 8. & Ericson Torleif (CERN) & Torleif.Ericson@cern.ch\\
 9. & Gasser J\"urg (Bern) & gasser@itp.unibe.ch\\
10. & Girlanda Luca  (Trento) & girlanda@ect.it\\
11. & Goldin, Daniel (CERN) & daniel.goldin@cern.ch\\
12. & Gotta Detlev (Juelich) & d.gotta@fz-juelich.de\\
13. & Guaraldo Carlo (Frascati) & guaraldo@lnf.infn.it\\
14. & Heim Thomas (Basel) & thomas.heim@unibas.ch\\
15. & Hencken Kai (Basel) & K.Hencken@unibas.ch\\
16. & Hirtl Albert (Wien) & albert.hirtl@oeaw.ac.at\\
17. & Karshenboim Savely (St.Petersburg) & sek@mpq.mpg.de\\
18. & Lamberto Antonino (Trieste) & Antonino.Lamberto@cern.ch\\
19. & Lanaro Armando (CERN) & Armando.Lanaro@cern.ch\\
20. & Leutwyler Heinrich (Bern) & leutwyler@itp.unibe.ch\\
21. & Lipartia Edisher (Lund) & lipartia@thep.lu.se\\
22. & Lyubovitskij Valery (Tuebingen) & valeri.lyubovitskij@uni-tuebingen.de\\
23. & Marel Gerard (Saclay) & marel@hep.saclay.cea.fr\\
24. & Moussallam Bachir (Orsay) & moussall@ipno.in2p3.fr\\
25. & Nemenov Leonid (Dubna) & Leonid.Nemenov@cern.ch\\
26. & Oades Geoffrey (Aarhus) & gco@phys.au.dk\\
27. & Pentia Mircea (Bucharest) & pentia@cern.ch\\
28. & Penzo Aldo (Trieste) & Aldo.Penzo@cern.ch\\
29. & Rappazzo Gaetana (Trieste) & Gaetana.Rappazzo@cern.ch\\
30. & Rasche Guenther (Zuerich) & rasche@physik.unizh.ch\\
31. & Sainio Mikko (Helsinki) & mikko.sainio@helsinki.fi\\
32. & Santamarina Cibran (CERN) & Cibran.Santamarina.Rios@cern.ch\\
33. & Sazdjian Hagop (Orsay) & sazdjian@ipno.in2p3.fr\\
34. & Schuetz Christian (CERN) & christian.schuetz@cern.ch\\
35. & Schweizer Julia (Bern) & schweizer@itp.unibe.ch\\
36. & Stern Jan (Orsay) & stern@ipno.in2p3.fr\\
37. & Trautmann Dirk (Basel) & Dirk.Trautmann@unibas.ch\\
38. & Voskresenskaya Olga (Dubna) & Olga.Voskresenskaja@mpi-hd.mpg.de\\
39. & Weissbach Florian  (Basel) & Florian.Weissbach@unibas.ch\\
40. & Yazkov Valery (Moscow) & Valeri.Iazkov@cern.ch\\
41. & Zemp Peter (Bern) & zemp@itp.unibe.ch\\
\end{tabular}

\newpage
\newcommand{\talk}[1]{\clearpage\label{#1}\input{#1}}                
\newcommand{\cont}[2]{\noindent #1\hfill\pageref{#2}\par\medskip}

\section{Scientific program}

\hfill Page
\medskip

\cont{{\bf L.~Girlanda}\\%
Need for new low-energy $\pi\pi$ scattering data}{girlanda}

\cont{{\bf J.~Stern}\\Chiral perturbation theory and massive strange quark pairs}{stern}

\cont{{\bf S.G.~Karshenboim}\\QED Theory of Hydrogen-like Atoms}{karshenboim}

\cont{{\bf J.~Gasser}\\Ground-state energy of pionic hydrogen to one loop}{gasser}

\cont{{\bf G.C.~Oades}, G.~Rasche and W.S.~Woolcock\\
The influence of the $\gamma n$ channel on the energy and width of pionic hydrogen}{oades}

\cont{{\bf T.A. Heim}, K. Hencken, M. Schumann, D. Trautmann and
  G. Baur\\Coupled channel approach to breakup of pionium}{heim}

\cont{{\bf C.~Santamarina} and L.G.~Afanasyev\\
On the influence of low n and high n states cross section
in the break-up probability\\ of pionium}{santamarina}

\cont{{\bf C. Guaraldo}\\Recent results from DEAR at DA$\Phi$NE}{guaraldo}

\cont{{\bf D.~Gotta} et al.\\
  The Pionic-Hydrogen Experiment at PSI}{gotta}

\cont{{\bf V.~Yazkov}\\Lifetime measurement of $\pi^+\pi^-$ atom at DIRAC}{yazkov}

\cont{{\bf C.~Cheshkov} and G.~Marel\\
Prospects for the study of the $K^\pm_{e4}$ decays at the NA48/2 experiment}{cheshkov}

\cont{{\bf L. Nemenov}\\Future experimental
  investigation of the $\pi^+\pi^-$ atom}{nemenov}

\cont{{\bf H. Sazdjian}\\Relevance of $K\pi$ atom measurements}{sazdjian}

\cont{{\bf P.~B\"uttiker}, S.~Descotes, and B.~Moussallam\\
The Roy equations for the ${\pi K}$ system}{buettiker}

\cont{{\bf S.~Descotes-Genon}, N.~Fuchs, L.Girlanda and J.~Stern\\
  $\pi\pi$ scattering and the chiral structures of QCD
  vacuum}{descotes}

\cont{{\bf A.Tarasov} and O.Voskresenskaya\\
  On the role of multi-photon exchanges in the incoherent interaction
  of $\pi^+\pi^-$-atom\\ with atoms of matter}{tarasov}

\cont{{\bf H.~Leutwyler}\\Concluding remarks}{leutwyler}

\newpage
\clearpage\label{girlanda}\begin{center}
{\Large{\bf Need for new low-energy $\pi\pi$ scattering data}}

\bigskip

{\bf L.~Girlanda\\[2mm]  
}
{\em ECT$^*$ and INFN, Strada delle Tabarelle 286, 38050 Trento, Italy}

\end{center}


The importance of low-energy $\pi\pi$ scattering data for probing the
mechanism of spontaneous breaking of chiral symmetry (SBChS) in QCD
has often been emphasized. Explicit calculations of the $\pi\pi$
scattering amplitude at two-loop level in the framework of chiral
perturbation theory (ChPT) allow to relate the scattering parameters
to the low-energy constants, describing the features of the
vacuum. Present knowledge about these constants, complemented with
standard assumptions about the size of the quark condensate (i.e. the
validity of the Gell-Mann--Oakes--Renner relation) and available
experimental information at medium and high energy, leads to very
precise predictions for the $S$-wave scattering lengths [1]
\begin{equation}
a_0^0=0.220 \pm 0.005, \quad a_0^2=-0.0444 \pm 0.0010.
\end{equation}
On the other hand, thanks to the solutions of the Roy Equations
provided by Ananthanarayan, Colangelo, Gasser and Leutwyler [2], 
the scattering amplitude is parametrized in terms of only two
parameters, the two $S$-wave
scattering lengths, within very small uncertainty in the whole energy
range below 800~MeV.
Therefore any experimental information below 800~MeV can in principle
be translated into a determination of the two $S$-wave scattering
lengths, thus providing a test of the predictions of ChPT. 
We applied this procedure [3] to the  recently published E865
data [4] on charged $K_{e4}$ decays, supplementing them with
available data in the $I=2$ channel below 800~MeV.
\begin{figure}[thb]
\begin{center}
\includegraphics[width=9.5cm,angle=270]{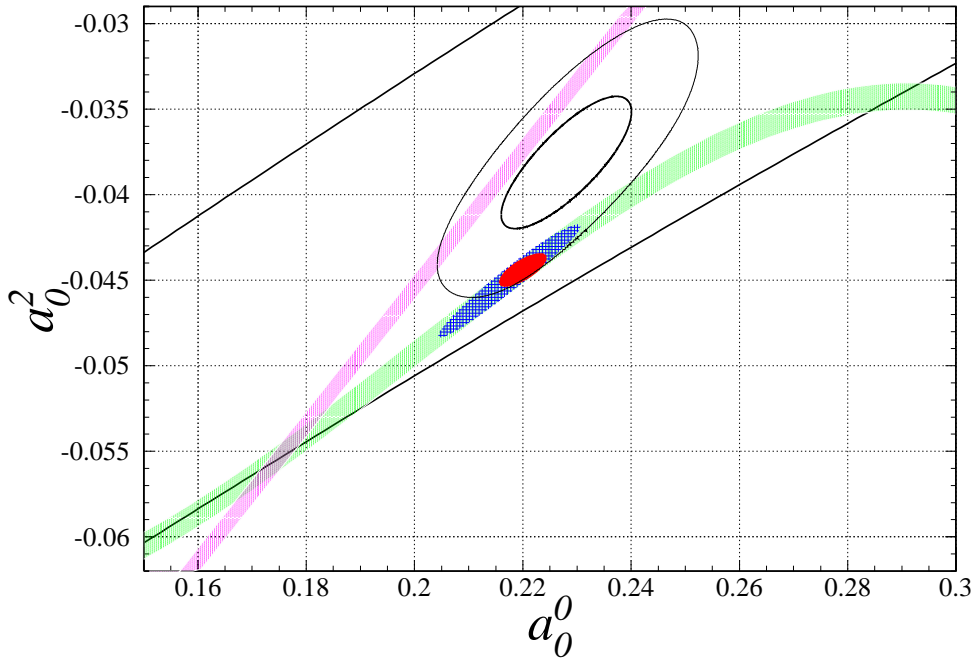}
\caption{Results of the two-parameter fit to the $K_{e4}$ data and
available $I=2$ data below 800~MeV (see the text). The two straight
lines delimit the region where Roy Equations admit solutions.}
\end{center}
\end{figure}

The ellipses in the figure (full lines) represent the 1-$\sigma$ and
2-$\sigma$ contours of our fit.  Our analysis is exclusively based on
the new solutions of the Roy Equations [2] and on
direct experimental information.
In particular no use is made of the correlation between $a_0^0$ and $a_0^2$
as inferred from the scalar radius of the pion [5]. If the dispersive
determination of the latter, $\langle r^2 \rangle_s = ( 0.61 \pm 0.04
)$~fm$^2$ is used, this correlation,
at the level of two-loop accuracy, is shown by the narrow curved band
of the figure.
The use of such a strong constraint would result in the filled shaded
ellipse, which is perfectly compatible with the two-loop ChPT
prediction, represented by the small filled ellipse. 
The two procedures yield different results at the 1-$\sigma$
significancy. However the narrow strip constraint depends
substantially on the NNLO contributions, which involve unknown $O(p^6)$
low-energy constants: the same correlation without the inclusion of
the $O(p^6)$ order would correspond to the narrow straight band in the figure.
In order to obtain the $O(p^6)$ narrow strip one has to rely  on
estimates of the constants appearing at this order, based on the
resonance saturation assumption. 
By reversing the argument, one can use the model-independent result
for the scalar scattering lengths
\begin{equation}
a_0^0=0.228 \pm 0.012,\quad a_0^2=-0.0382\pm0.0038,
\end{equation}
to test the narrow strip prediction, i.e. to determine the $O(p^6)$
 constants. The latter are found unexpectedly 
 large, which can be interpreted as a manifestation of the exceptional
 status of the  scalar channel, characterized by a strong $\pi\pi$
 continuum and OZI rule violation. 
 The situation will hopefully be further clarified by the awaited
 results of DIRAC and the forthcoming $K_{e4}$ experiment NA48-II.
 Unfortunately it is unlikely that the measurement of the pionium
 lifetime can distinguish between the two procedures, since all the
 ellipses shown in the figure correspond to very similar values for
 the combination $a_0^0-a_0^2$.

The fit results for $a_0^0$ and $a_0^2$ can be used to extract the
 values  of the subthreshold parameters $\alpha$ and $\beta$, of the
low-energy constants $\bar l_3$ and $\bar l_4$ as well as of the main
two-flavour order parameters: $\langle \bar u u \rangle$ and $F_{\pi}$ in
the limit $m_u=m_d=0$ taken at the physical value of the strange quark mass.
The results for the order parameters read
\begin{equation}
X(2)=\frac{2 \hat m |\langle \bar q q \rangle |}{F_{\pi}^2 M_{\pi}^2}
= 0.81 \pm 0.09, \quad Z(2)=\frac{F^2}{F_\pi^2} = 0.90 \pm 0.03,
\end{equation}
signalling that the two-flavour standard ChPT is the appropriate
expansion scheme.


\clearpage\label{stern}\begin{center}
{\Large{\bf Chiral perturbation theory and massive strange quark pairs}}

\bigskip

{\bf Jan Stern }\\[2mm]  

 {\em Groupe de Physique Theorique, Institut de Physique Nucleaire,
          Universite de Paris- Sud , 91406 ORSAY - CEDEX }
\end{center}

Low - energy $\pi\pi$ scattering merely involves valence $ u$ and $d$ quarks
and it probes $QCD$ in the vicinity of the $SU(2)\times SU(2)$ chiral limit
in which $ m_u = m_d = 0$ and $m_s$ is fixed at its physical value.
Nevertheless, the influence of massive virtual $\bar ss$ pairs on low-energy
$\pi\pi$ scattering, on the two-flavour chiral order parameters
$\Sigma(2) = -\lim_{m_u,m_d\to 0}\langle \bar uu \rangle|_{m_s=phys}=F^2B$,
$F^2 = \lim_{m_u,m_d \to 0} F^2_{\pi}$ , as well as their impact
on the estimates of
low energy constants $\bar l_3$ and $\bar l_4$  [1] may be important for the
following two reasons:  i) First,$m_s\ll \Lambda_H\sim 4\pi F_{\pi}\sim 1GeV$  
suggesting some overall
convergence of the $SU(3) \times SU(3)$ chiral expansion . On the other
hand, as long as $m_s\sim\Lambda_{QCD}$, the scale at which
$QCD$ becomes strong, the vacuum gets populated by abundant virtual
$\bar ss$ pairs, whereas the creation of heavy quark pairs is suppressed.
ii) The second reason is the existence of a strong long range correlation
between strange and non-strange $\bar qq$ pairs in the $J^P=0^+$ channel:
\begin{equation}\label{Z}
Z^s = \lim _{m_u,m_d\to 0} i\int dx\langle{T\bar ss(x)\bar uu(0)}\rangle_c 
\end{equation}
is subleading in large $N_c$ expansion and it is  expected to be suppressed
by the OZI - rule. Instead, recent sum rule estimates [2] suggest that
 $Z^s$ is of a normal size
$Z^s \sim \Lambda^2_{QCD}$ , reflecting the violation of the OZI-rule and
of the $1/N_c$ expansion observed in the scalar channel. This phenomenological
information was not available in the past and most of the original $\chi$PT
considerations and estimates  [1], (including the G$\chi$PT variant)
were based on the OZI-rule and large $N_c$ wisdom.  Today, the
strong vacuum OZI-rule violating correlation $Z^s$ provides a new element
in the discussion and it naturally fits to our microscopic understanding
of $\chi$SB in Euclidean QCD: The quark condensate represents
{\bf the average} and the correlator $Z^s$  {\bf the fluctuations} of the
density of smallest modes of the Dirac operator.Notice that $Z^s>0$.

            Vacuum fluctuations of $\bar ss$ pairs may lead to important
$m_s$ dependence of the two-flavour condensate $\Sigma(2)$.  At $m_s=0$,
$\Sigma(2)$ coincides with the three-flavour condensate
$\Sigma(3) = - \langle \bar uu \rangle|_{m_u=m_d=m_s=0} = F^2_0 B_0 $
and its derivatives
with respect to $m_s$ is precisely the correlator $Z^s$ (related to the low
- energy constant $L_6$). Up to higher order terms one has
\begin{equation}\label{induced}
\Sigma(2) = \Sigma(3) +  m_s Z^s + \ldots   .
\end{equation}
$\Sigma(3)$ is the ``genuine condensate'' of $QCD$ with three massless quarks
and no massive quarks left which would be enough light to pollute the vacuum.
$m_sZ^s$ represents the positive contribution to the two-flavour
condensate $\Sigma(2)$
that is induced from massive $\bar ss$ vacuum pairs. The two contributions
can be of sensibly comparable size [2], reflecting the suppression of
$\Sigma(3)$ relative to $\Sigma(2)$  due to vacuum fluctuations.
This can be seen from a non-perturbative analysis of Goldstone-boson mass
and decay constant Ward identities [3], provided the quark mass ratio
$r = m_s/m$ is not too small ($r>15$). For small $Z^s$, which in turn
implies a precise fine tuning of $L_6(M_\rho)$ to the critical value
$L^{crit}_6 = - 0.26 \times 10^{-3}$ , one would have
$\Sigma(2) \approx \Sigma(3)$. For $L_6$ slightly above $L^{crit}_6$,
the fluctuations grow and the yield of the genuine
condensate $\Sigma(3)$ in Eq.~(\ref{induced})decreases. This decrease is,
however, compensated by a growing induced contribution $m_sZ^s$. As a result,
the massive $\bar ss$ pairs stabilize the two-flavour condensate $\Sigma(2)$
and keep it large independently of vacuum fluctuations and of the
suppression of the genuine condensate $\Sigma(3)$ ,(see the talk by
S. Descotes-Genon).

       It is instructive to compare  the $N_f=2$ $\chi$PT
(i.e. expansion in powers
of   $m_u=m_d=m$ for a fixed $m_s$) with the $N_f=3$ $\chi$PT expansion
(in powers of $m$ {\bf and} $m_s$) of the same quantity $F^2_\pi M^2_\pi$.
Keeping the LO and the NNLO contributions, one gets
\begin{equation}\label{two}
F^2_{\pi}M^2_{\pi}=2m\Sigma(2)+\frac{m^2B^2}{8\pi^2} (4\bar l_4 -\bar l_3)+\ldots 
\end{equation}
\begin{equation}\label{three}
F^2_{\pi}M^2_{\pi}=2m\Sigma(3) + 2mm_s Z^s + 4 m^2 (Z^s + A ) +\ldots .
\end{equation}
respectively. The two expansions have to coincide order by order in m.
The most important part of the OZI - rule violating contribution to
rhs of Eq.~(\ref{three}), $m_sZ^s$ , is absorbed into $\Sigma(2)$.
The contribution of $Z^s$ to the LEC's $\bar l_3 ,\bar l_4$ is not
enhanced by the factor $m_s$ and it should be of a comparable size
as the ``normal'' contribution $A$ (related to $L_8$). Hence, vacuum
fluctuations will slightly shift
the OZI-rule based estimate [1] $\bar l_3 = 2.9 \pm 2.4$     towards a
negative values  without spoiling the convergence
of the  $N_f=2$ $\chi$PT. The $N_f=2$ expansion is dominated by the
condensate term $\Sigma(2)$, since the latter is enhanced by the
contribution induced by $\bar ss$ pairs. On the other hand, the genuine
condensate $\Sigma(3)$ need not dominate the $N_f=3$ expansion of Goldstone
boson masses, since the vacuum fluctuation term $m_s Z^s$  now counts as NLO.
It is useful to express both condensates in  appropriate GOR units
\begin{equation}\label{X}
X(N_f) = \frac{2m\Sigma(N_f)}{F^2_{\pi}M^2_{\pi}} \qquad N_f = 2 , 3
\end{equation}
and to compare $X(2) > X(3)$ with one. The $N_f=2$ low-energy parameters can
be extracted from precise $\pi\pi$ scattering experiments .      
Our analysis [4] including
most recent experimental data yields $X(2)=0.81\pm 0.07$ ,
$F^2/F^2_\pi = 0.89\pm 0.02$ and $ \bar l_3 = - 17 \pm 15$ confirming
both the dominance of the two-flavour condensate and the expected shift
of $\bar l_3$. (See the talk by L. Girlanda for more details.)

                Suppression of $X(3)$ does not necessarily mean a bad
convergence of the $N_f=3$ $\chi$PT for ``good observables'' such as
(\ref{three}). However, provided in the latter the LO
term $2m\Sigma(3)$ and the NLO term $2mm_sZ^s$ are of a comparable size,
we face an instability which prevents us to proceed perturbatively eliminating
low-energy parameters in terms of observables. A systematic non-perturbative
alternative to this standard procedure does exist and it makes a new $N_f=3$
analysis of precise $\pi\pi$ , $\pi$K and $\eta$-decay data both meaningful
and interesting.


\clearpage\label{karshenboim}\begin{center}
{\Large{\bf QED Theory of Hydrogen-like Atoms}}
\bigskip

{\bf Savely G. Karshenboim}\\[2mm]  
{\em D. I. Mendeleev Institute for Metrology (VNIIM), St. Petersburg 198005, Russia}
{\em Max-Planck-Institut f\"ur Quantenoptik, 85748 Garching, Germany}
\end{center}

About fifty years ago the experimental discovery of the Lamb shift and the anomalous 
magnetic moment of electron initiated a long competition between 
QED theory and experiment. Today a list of quantities used to test QED 
is quite impressive and includes:
\begin{itemize}
\item the anomalous magnetic moments of the electron and muon;
\item hyperfine splitting of the $1s$ and $2s$ states in hydrogen and 
deuterium atoms and helium-3 ion;
\item gross and fine structure intervals and the Lamb shift in hydrogen, 
deuterium and helium-4 ion;
\item  hyperfine splitting and $1s-2s$ transition in muonium and 
positronium,
\item decay rates of para- and orthopositronium including rare modes;
\item energy levels of high- (like e.g. bismuth) and medium-Z 
(like e.g. carbon) ions with few electrons;
\item $g$ factors of a bound electron in hydrogen-like atoms;
\item energy levels of muonic and exotic atoms;
\item energy levels of three-body atoms (helium, antiprotonic helium, 
muonic helium).
\end{itemize}

Twenty years ago the experimental accuracy of most QED tests was 
significantly better than the theoretical one. With time situation changed
so that after successful devopments modern QED theory dominates 
over experiment.

However, there is the third side in this competition. Unfortunately,
theory and experiment cannot speak the same language. Experiment leads
to results expressed in units of Hertz and Mev, while theory delivers
only expressions containing values of fundamental constants (such as
the fine structure constant $\alpha$, electron mass $m_e$) and
particle/nuclear parameters (like e.g. the proton charge radius). In
other words, theory is actually not in position to predict any numbers
building instead bridges between different experiments. As a bridge,
theory is the most accurate part of most QED tests. However, trying to
calculate some quantity needed to make a prediction we often discover
that the input data of the calculation, i.e. the fundamental and
auxiliary constants, are not known accurately enough from their
measurements in other experiments.

In the case of hadronic effects for some QED quantities (like e.g. 
hyperfine interval in the hydrogen atom or anomalous magnetic moment of muon) 
we even need functions (such as the proton electric form factor) to be 
somehow determined as the input data.

The lesson we have to learn from precision studies of simple atoms and 
QED tests is that the situation for free particles ($g-2$ for electron 
and muon), low $Z$ atoms (hydrogen, muonium), high $Z$ atom (bismuth, lead), 
muonic and exotic atoms is actually quite similar:
\begin{itemize}
\item the QED calculations themselves can be done with high enough
  accuracy to compete with experiment. Actually, QED theory rather
  wins the competition.
\item However, that is not enough:
\begin{itemize}
\item we need to have access to accurate values of the fundamental
  constants to transform QED expressions into numbers;
\item we also need to know some functions like form factors to put
  into the calculation.
\end{itemize}
\item Eventually, comparison of theory and experiment is often not
  quite a QED test, but rather a determination of some fundamental
  constants or study of nuclear effects.
\end{itemize}

The crucial orders of magnitude important for comparison of theory and 
experiment are summarized in the Table below. In most of the cases the 
QED tests confirm theory. 

\begin{center}
\begin{tabular}{lc}
\hline\hline
Quantity & Order \\
\hline 
Hydrogen, deuterium (gross structure) & $\alpha(Z\alpha)^7m$, $\alpha^2(Z\alpha)^6m$ \\
Hydrogen (fine structure and Lamb shift) & $\alpha(Z\alpha)^7m$, $\alpha^2(Z\alpha)^6m$ \\
$^3$He$^+$ ion ($1s/2s$ hyperfine structure)  & 
$\alpha(Z\alpha)^7m^2/M$,$\alpha(Z\alpha)^6m^3/M^2$,\\
& $\alpha^2(Z\alpha)^6m^2/M$, $(Z\alpha)^7m^3/M^2$\\
$^4$He$^+$ ion (gross structure and Lamb shift)  & 
$\alpha(Z\alpha)^7m$, $\alpha^2(Z\alpha)^6m$ \\
Muonium ($1s$ hyperfine structure)      & 
$(Z\alpha)^7m^3/M^2$, $\alpha(Z\alpha)^6m^3/M^2$,\\  
&$\alpha(Z\alpha)^7m^2/M$ \\
Positronium (gross structure and $1s$ hyperfine structure)  & $\alpha^7m$ \\
Parapositronium (decay rate)       & $\alpha^7m$ \\
Orthopositronium (decay rate)      & $\alpha^8m$ \\
Parapositronium ($4\gamma$ branching)  & $\alpha^8m$ \\
Orthopositronium ($5\gamma$ branching) & $\alpha^8m$ \\
\hline\hline
\end{tabular}
\end{center}
\vspace{-0.2cm}
\centerline{
\em 
Table 1. Comparison of QED theory and experiment: crucial orders of magnitude\bigskip}

More detail on comparison of theory versus experiment can be found in [1]. 
A detailed review of theory is presented in [2]. Most of recent results 
and projects related to QED tests and simple atoms along with reviews were 
presented at two conferences on {\em Precision Physics of Simple Atomic 
Systems} (PSAS) in 2000 (Castiglione della Pescaia) [3] and 2002 
(St. Petersburg) [4].

This work was supported in part by the RFBR grant 00-02-16718.


\clearpage\label{gasser}\begin{center}
{\Large{\bf Ground-state energy of pionic hydrogen to one loop}}

\bigskip

{\bf J.~Gasser}\\[2mm]  

{\em Institute for Theoretical Physics, University of Bern, Sidlerstrasse
  5, CH-3012 Bern}
\end{center}

In my talk I discussed recent work on pionic hydrogen performed in Ref.~[1].
In Ref.~[2], the relation between the scattering length 
combination $|a_{0+}^+ - a_{0+}^-|$ and the strong energy-level shift 
of pionic ground state has been worked out at leading order in the 
low-energy expansion of the isospin breaking correction.
In Ref. [1], we have  
carried out the calculation at next-to-leading order, where the effect of 
 the photon-neutron intermediate state first shows up in the chiral expansion.
 This 
amounts to evaluating the elastic $\pi^-p \rightarrow \pi^- p$ amplitude 
at order $p^3$, including isospin breaking contributions 
generated by the quark mass difference
 $m_u-m_d$ and by virtual photon loops. In order to have a systematic
 framework, we have extended the infrared regularization method 
developed by Becher and Leutwyler~[3], such that it includes virtual 
photons as well. Furthermore, we have performed a heat-kernel evaluation of 
all singular parts in the one-loop calculation, including meson, 
baryon and photon fields. As far as we are aware, this provides for the 
first time a consistent set of effective lagrangians at 
this order in the low-energy expansion. We find 
that triangle-type graphs generate large corrections, whereas
 the effect of the low-energy constants at order $p^3$ is
 suppressed by the factor $m_\pi/m_{\mathrm{proton}}$. 
On the other hand, 
 the  constant $f_1$, that occurs  at leading order,  may contribute 
 significantly to the energy shift~[1]. Unfortunately, 
 it has not yet been determined  with sufficient accuracy.
 We conclude that the potential-model calculations available in the 
 literature largely underestimate
 the systematic error, because these models are not properly 
 matched to the underlying theory (QCD+QED). Furthermore, claims made in the 
literature that
 a precise measurement of the strong energy-level shift in 
pionic hydrogen will
 provide a corresponding precise determination of 
 $|a_{0+}^+ - a_{0+}^-|$ are too optimistic 
at the present stage of the theoretical knowledge: 
Experiment is ahead of theory in this case. 
For further references and details, I refer to [1]. 
A very recent article is [4] - we plan to comment on  this work elsewhere [5].


\clearpage\label{oades}
\begin{center}
{\Large \textbf{The influence of the $\gamma n$ channel on the energy and
width of pionic hydrogen}}

\bigskip

\textbf{G.C.~Oades$^1$, G.~Rasche$^{2}$ and W.S.~Woolcock$^3$}\\[2mm]

$^1$ \emph{Institute of Physics and Astronomy, Aarhus University, DK-8000
Aarhus C, Denmark.}

$^2$ \emph{Institute f\"{u}r Theoretische Physik der Universit\"{a}t,
Winterthurerstrasse 190, CH-8057 Z\"{u}rich, Switzerland.}

$^3$ \emph{Department of Theoretical Physics, IAS, The Australian National
University, Canberra, ACT 0200, Australia.}
\end{center}

We reconsider the corrections presented in the paper of Sigg et al. [1] for
the extraction of the hadronic $s-$wave $\pi N$ scattering lengths from
measurements of the energy and width of pionic hydrogen. Our starting point
is the Deser type formulae given in the multi-channel situation by Rasche and
Woolcock [2]. These relate the energy shift to the nuclear scattering
length, $a_{cc}^{n}$, for the process $\pi ^{-}p\rightarrow \pi ^{-}p$ and
the width for decay to the $\pi ^{0}n$ channel to the scattering length, $%
a_{c0}^{n}$ for the process $\pi ^{-}p\rightarrow \pi ^{0}n$. These nuclear
scattering lengths are in turn related to the hadronic scattering lengths by
the expressions 
\begin{eqnarray*}
a_{cc}^{n}=a_{cc}^{h}(1+\delta _{cc}),
\\
a_{c0}^{n}=a_{c0}^{h}(1+\delta _{c0}),
\end{eqnarray*}
where $\delta _{cc}$ and $\delta _{c0}$ are the electromagnetic corrections. Values
for these were first considered in [2], later examined in more detail in [1]
and more recently recalculated in [3].

In addition to the hadronic channels $\pi ^{-}p$ and $\pi ^{0}n,$ pionic
hydrogen also decays to the channel $\gamma n,$ the widths being related by
the Panofsky ratio
\[
\frac{\Gamma _{\pi ^{0}n}}{\Gamma _{\gamma n}}=P=1.546\pm 0.009.
\]
This channel also influences the EM corrections and the first crude
estimates of its effect were made in [1] where the effective potential in
the $\pi ^{-}p\rightarrow \pi ^{0}n$ channel was empirically increased so
that it gave a value $a_{c0}^{eff}$ which reproduced the full width i.e.
\[
a_{c0}^{eff}=a_{c0}^{n}\sqrt{1+P^{-1}}.
\]
This modified potential was then used to calculate new EM corrections
and the differences between these values and the values from the unmodified
2-channel calculation, $\Delta \delta _{cc}$ and $\Delta \delta _{c0}$,
were assigned to the influence of the $\gamma n$ channel. The final values
with errors given in [1] were
\begin{eqnarray*}
\delta _{cc} &=&-(2.1\pm 0.5)\% \\
\delta _{c0} &=&-(1.3\pm 0.5)\%
\end{eqnarray*}
where the contributions from the $\gamma n$ channel were
\begin{eqnarray*}
\Delta \delta _{cc} &=&-0.73\% \\
\Delta \delta _{c0} &=&+0.15\%.
\end{eqnarray*}

What the authors of [1] overlooked was that the modification of the
effective potential in the $\pi ^{-}p\rightarrow \pi ^{0}n$ channel not only
changes $a_{c0}^{n}$ to $a_{c0}^{eff}$ but also modifies $a_{cc}^{n}$, an
effect which can also be considered as a type of EM correction. In
order to examine this effect we have extended the calculations used in [3]
to the case where we have the three channels, $\pi ^{-}p$, $\pi ^{0}n$ and $%
\gamma n.$ Such calculations involving mass $0$ photons are possibly
dangerous due to the fact that the reduced mass of the $\gamma n$ system is $%
0$. As pointed out in [4], this problem does not occur in the case of a
relativistic equation of motion. In our case we use the relativised Schr\"{o}%
dinger equation where the appropriate term in the non-relativistic equation, 
$2M_{\gamma n}^{red}V_{c\gamma }$, becomes $2M_{\gamma n}^{red}f_{c\gamma
}V_{c\gamma }$ where
\[
M_{\gamma n}^{red}f_{c\gamma }=\frac{M_{n}m_{\gamma }}{M_{n}+m_{\gamma }}%
\frac{W^{2}-M_{n}^{2}-m_{\gamma }^{2}}{2m_{\gamma }W}
\]
which has a finite limit as $m_{\gamma }\rightarrow 0$. In this calculation
we also need effective potentials to describe the three channel system. For
the $\pi ^{-}p$ and $\pi ^{0}n$ channels we use hadronic potentials fitted
to the low energy phase shifts together with the Coulomb and vacuum
polarization potentials corrected for finite size effects. For $\pi
^{-}p\rightarrow \gamma n$ and $\pi ^{0}n\rightarrow \gamma n$ we adjust the
potentials to fit the $E_{0+}$ photoproduction multipole amplitudes and for $%
\gamma n\rightarrow \gamma n$ we adjust to fit the low energy $\gamma
n\rightarrow \gamma n$ cross section. We now make the same type of
calculations as in the two channel case i.e. we compare the value of the
threshold $K$-matrix calculated in the full three channel case with its
purely hadronic value. For the EM corrections, preliminary calculations show
only a very small change in $\delta _{c0}$ but $\delta _{cc}$ increases to a value around zero. 

Can we understand this physically? What we are seeing might well be called
an extended Ball-Frazer mechanism. The addition of the extra inelastic
channel gives an effective attraction which increases the binding energy.
This seems to be an effect which comes from the requirements of multichannel
unitarity and as such will not be seen in lowest order calculations which
ignore these requirements.

Another very important point is that our EM corrections refer to a
hypothetical hadronic world which is, by tradition, defined as one in which
nucleons have the mass of the proton, pions the mass of the charged pion
and where the effective potentials are isopin invariant. This should be
remembered when comparing to the results of Gasser et al. [5] who refer
to a hadronic situation consistent with chiral perturbation theory. We could,
of course, use a similar definition of our hadronic reference point. The
hadronic masses would then be taken from chiral perturbation theory and
the effective potentials adjusted to reproduce the low energy chiral 
perturbation theory amplitudes. To do this without assuming isospin invariance
requires the low energy amplitudes in all three channels,
$\pi ^{-}p\rightarrow \pi ^{-}p$, $\pi ^{-}p\rightarrow \pi ^{0}n$ and
$\pi ^{0}n\rightarrow \pi ^{0}n$ and we therefore ask the chiral perturbation 
theory experts for these values so that we can proceed with this work. 


\clearpage\label{heim}\begin{center}
{\Large\bfseries Coupled channel approach to breakup of pionium}

\bigskip

{\bfseries \underline{T.A. Heim}$^1$, K. Hencken$^1$, 
M. Schumann$^1$, D. Trautmann$^1$ 
and G. Baur$^2$}\\[2mm]  

$^1$ \emph{Institut f\"ur Physik, Universit\"at Basel, 
Klingelbergstrasse 82, CH--4056, Basel, Switzerland}

$^2$ \emph{Institut f\"ur Kernphysik, Forschungszentrum J\"ulich, 
52425 J\"ulich, Germany}
\end{center}
Since the experiment DIRAC does not directly observe the neutral pions 
from the strong decay channel $\pi^+\pi^-\rightarrow 2\pi^0$, but rather 
looks at the charged pions from electromagnetic breakup in the target, 
this electromagnetic breakup process must be calculated with very high
accuracy (on the order of 1\%) as a required input for the analysis
of the experiment. In order to extract the pionium lifetime from
the observed breakup probability, the passage of pionium through the
target matter is studied in a simulation which in turn requires detailed
knowledge of the cross sections for bound-bound and bound-free transitions
of the pionium [1].
 
In previous work [2--4] we have shown that the semiclassical approximation
provides the appropriate tool for this problem. In first order Born
approximation the accurate description of both the pionium and the target
system with its atomic structure, as well as their relativistically correct
interaction (including magnetic terms), has been implemented
successfully (accurate to far better than 1\% within this model). 
However, it has been noted by studying the total cross sections [5],
and veryfied by explicitly calculating the bound-bound transitions in
the Glauber formalism [6] that higher order contributions can change the
first order results by up to several per cent. We have estimated the validity
of the Glauber approximation in [6] by studying the influence of finite
interaction times within a simple extension of the 
sudden approximation. We found that the results varied
within less than 0.1\% which may thus be regarded as the accuracy of the
higher order calculation (clearly below the required 1\% limit). 

\noindent
\begin{minipage}{8.5cm}
\mbox{}

In order to provide a further, independent test of the validity of the 
Glauber approximation, we study the time evolution of the occupation
probabilities for a few selected bound states of the pionium in a coupled 
channel approach. The figure shows this   
time evolution for various channels as 
indicated. In this example, the pionium starts out initially in
the 3s-state  (all other states not occupied; Nickel target). 
As time evolves,
other pionium states acquire non-zero occupation probabilities. 
Our goal is to verify whether the cross sections obtained in this approach
agree with our previous (numerically less demanding) calculations in 
a few test cases.
\end{minipage}
\hfill
\begin{minipage}{7.5cm}
\includegraphics[width=7.5cm]{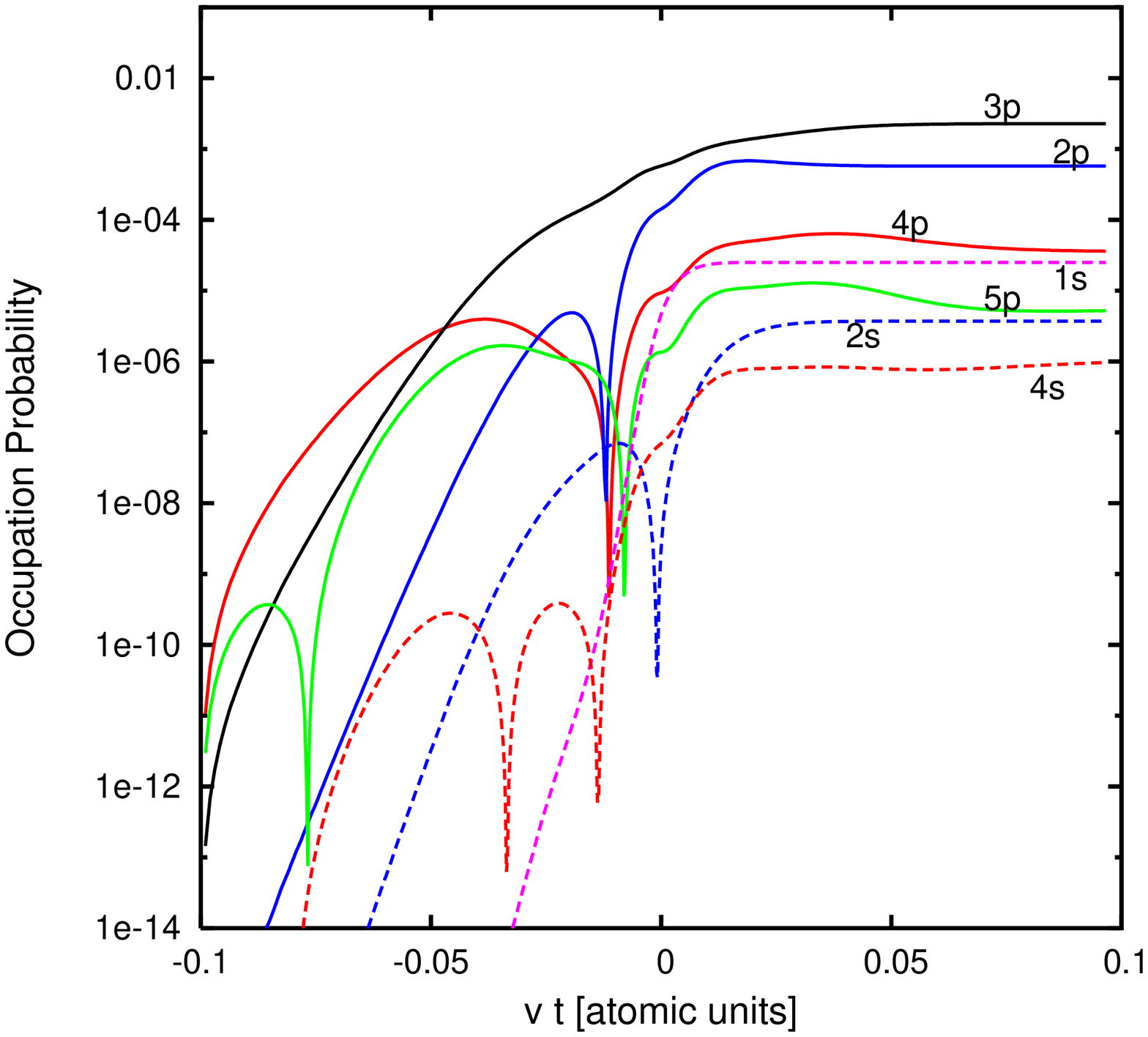}
\end{minipage}

\clearpage\label{santamarina}\begin{center}
{\Large{\bf On the influence of low n and high n states cross section
in the break-up probability of pionium}}

\bigskip

{\bf \underline{C.~Santamarina}$^1$ and L.G.~Afanasyev$^2$}\\[2mm]  


$^1$ {\em Institute of Physics, University of Basel, CH-4056, Basel
Switzerland}

$^2$ {\em Joint Institute for Nuclear Research, Dubna, Moscow Region, 141980 Russia}

\end{center}


The study of the break-up probability ($P_{br}$) dependence
of pionium as a function
of lifetime in a target of definite
material and thickness is a master key in the measurement that
DIRAC collaboration aims to achieve in its experiment PS-212 at CERN [1].

A detailed explanation of the physical problem can be found in [2] and
[3] and can be resumed as to solve the differential equation system that
links the population $p_{nlm}$ of the different atomic bound states
(defined by the
usual quantum numbers of an hydrogen-like system, $n$, $l$ and $m$)
as a function of the target position of the atom ($s$):
\[
\frac{dp_{nlm}(s)}{ds} = 
\sum_{n^{\prime} l^{\prime} m^{\prime}} 
a^{n^{\prime} l^{\prime} m^{\prime}}_{n l m} 
p_{n^{\prime} l^{\prime} m^{\prime}}(s) \qquad
\left\{ 
\begin{array}{l}
a^{n^{\prime} l^{\prime} m^{\prime}}_{n l m} = 
\frac{\sigma^{n^{\prime} l^{\prime} m^{\prime}}_{nlm} \rho N_0}{A} \\
a^{n l m}_{n l m} = \frac{\sigma^{total}_{nlm} \rho N_0}{A}- 
\left\{
\begin{array}{ll}
\frac{2 M_{\pi}}{P c \tau_{n}} & \mbox{if $l=0$}\\
0 & \mbox{if $l \neq 0$}
\end{array}
\right.
\end{array}
\right.
\]
where $\rho$ is the density and $A$ the atomic weight of the target, $N_0$
the Avogadro number, $M_{\pi}$ the mass of the pion, $\tau_{n}$ the lifetime
of the bound state and $P$ the center of mass momentum of the atom (average
around 4.2 $GeV/c$ in DIRAC experimental conditions). However, the main
input to the equation is given by the pionium-target atom interaction
cross sections, either for the probability of a transition between
bound states ($\sigma^{n^{\prime} l^{\prime} m^{\prime}}_{nlm}$) or for
the total interaction probability ($\sigma^{total}_{nlm}$).

These cross sections have been calculated in the Born approximation
using different parameterizations for the atomic form factors [2,4] and
with the Glauber [5] approximation that accounts for multi-photon exchange.
The three calculations lead to similar results with discrepancies between
$1\%$ and $4\%$ if the states involved in the transition are low
energy ones ($n \leq 5$) and increase as $n$ does up to differences
higher than $10\%$.

However, the difference in the breakup probability result of pionium is
to the level of $1\%$. As an example, asumming $2.91$ femtoseconds for
pionium lifetime: 
\[
P^G_{br} = 0.454 \qquad P^{B1}_{br} = 0.460 \qquad P^{B2}_{br} = 0.465
\]
where $P^G_{br}$ was calculated with the Glauber cross sections of [5],
$P^{B1}_{br}$ with the Born cross sections of [2] and $P^{B2}_{br}$
with the Bron cross sections of [4].

This agreement between the three results is explained by four things:
\begin{enumerate}
\item The atomic bound states are produced (initial conditions)
according to: $p^{prod}_n
\propto 1/n^3$ [6] and hence the amount of atoms created with
$n\geq 5$ is negligible.
\item The probability of annihilation also behaves as $p^{annih}_n
\propto 1/n^3$ [7] and hence there are almost no atoms annihilated from
states with $n\geq 5$.
\item The atoms show a clear tendency to be excited or broken rather
than to be de-excited, $\sum_{n^{\prime}< n, l^{\prime}, m^{\prime}}
\sigma^{n^{\prime} l^{\prime} m^{\prime}}_{nlm} \gg
\sum_{n^{\prime}\geq n, l^{\prime}, m^{\prime}}
\sigma^{n^{\prime} l^{\prime} m^{\prime}}_{nlm}$ (see Figure).
\item The total cross section increases very fast as $n$ does. And as
a consequence the mean free path of the atom ($\lambda_{nlm} =
A/\sigma^{total}_{nlm} \rho N_0$) decreases very fast with $n$ (see Figure).
\end{enumerate}

All this means that the states with $n \geq 5$ are populated 
by excitations from lower $n$ states since very rarely atoms are directly
created in these states. Moreover, de-excitations from these states to
lower $n$ states very seldom happen also. This means that the solution
of the differential equation system for the low $n$ states (where
annihilation probability is not negligible) does not
depend on the population of high $n$ states and hence, the discrepancies
between the different sets of cross sections will not lead to differences
in the solutions of the equation system for these states populations.
Meanwhile, an atom which
is excited into a high $n$ state has a very small free path (in all the
three sets of cross sections) compared
to the target dimensions of DIRAC (tenths of micrometers) and will
interact with very high probability until it ends up broken. The discrepancies
between the different sets of cross sections will be then irrelevant since
they all lead to a breakup probability result of  practically $100\%$ for
a bound state with $n \geq 5 $~\footnote{However this would not be the case
if the target thickness was much smaller.}.

\begin{figure}[thb]
\begin{center}
\includegraphics[width=5cm]{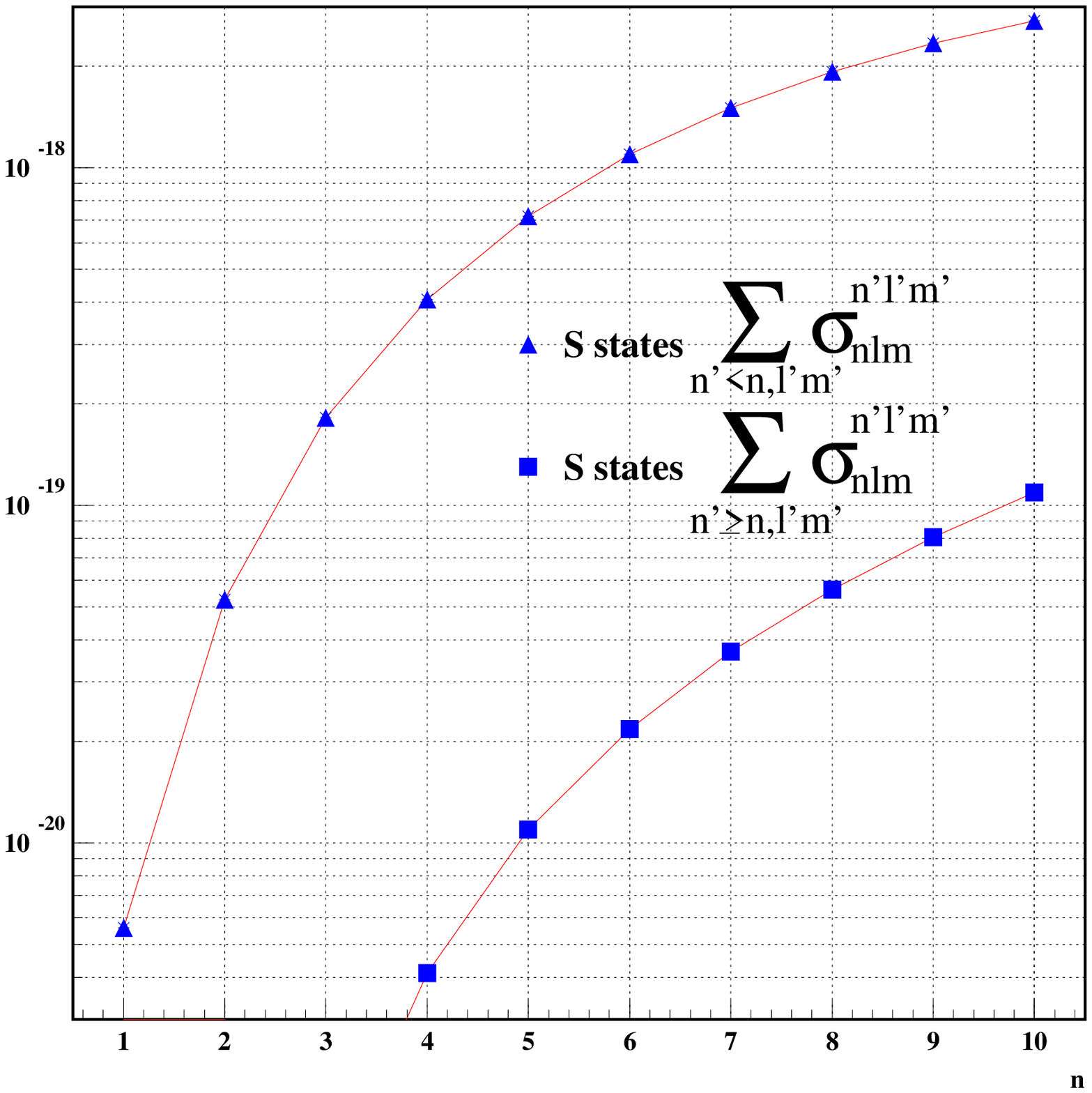}
\includegraphics[width=5cm]{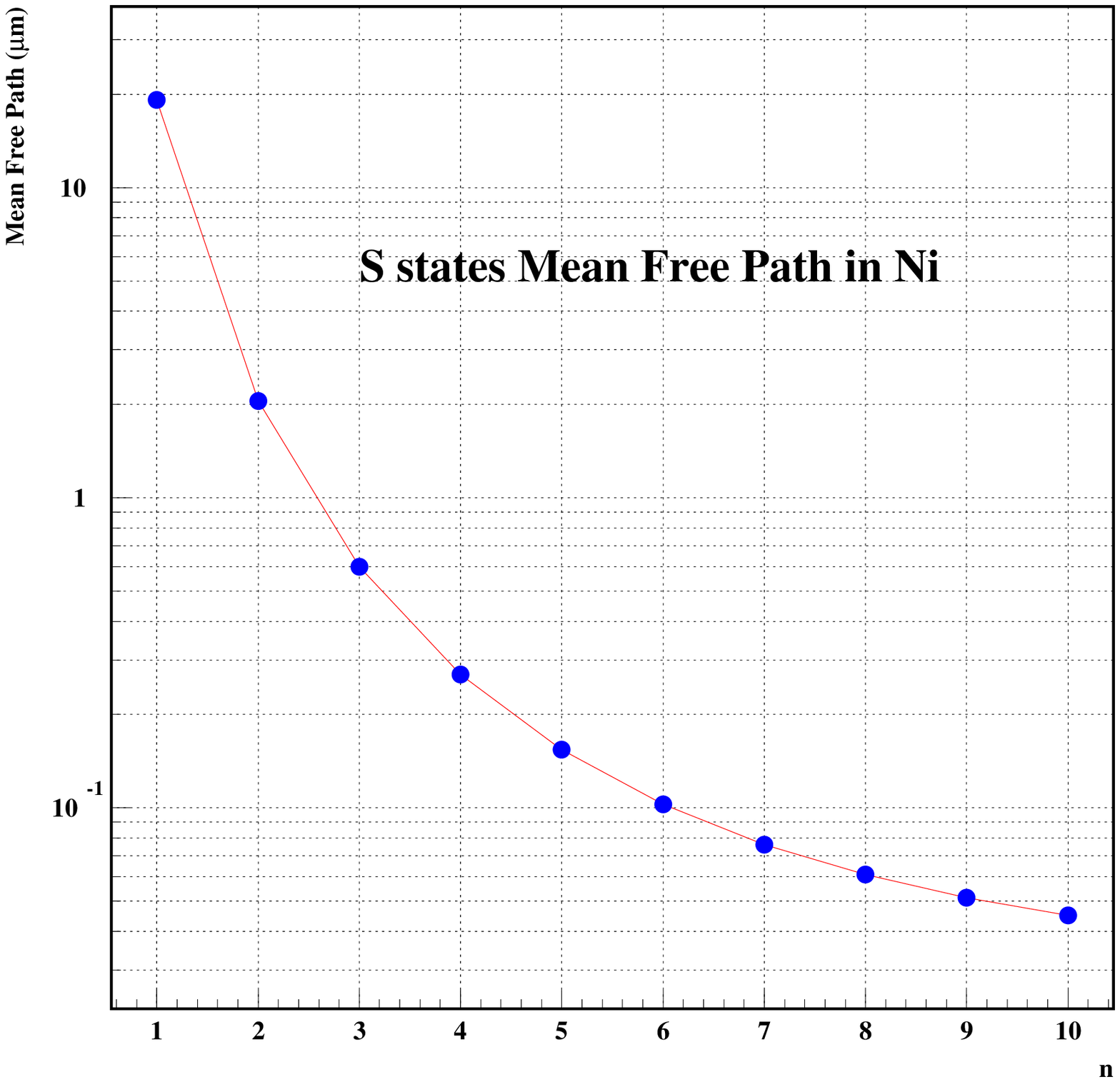}
\caption{On the left we show up the comparison between the excitation and the
de-excitation cross sections of [2] for the S sates of the atoms. On the
right we can observe the mean free path value of S states.}
\end{center}
\end{figure}

We want to thank Prof. Dr. Dirk Trautmann and his group for have kindly 
provided us their cross sections sets.


\clearpage\label{guaraldo}\begin{center}
{\Large{\bf Recent results from DEAR at DA$\Phi$NE}}

\bigskip

{\bf C. Guaraldo} on behalf of the DEAR Collaboration\\[2mm]

{\em  LNF-INFN, Via E. Fermi 40, 00044 Frascati (Roma), Italy }

\end{center}

In May 2001 the DEAR (DA$\Phi$NE Exotic Atom Research) collaboration [1]
performed the first measurement of an exotic atom (kaonic nitrogen) at the
DA$\Phi$NE collider of the Laboratori Nazionali di Frascati dell'INFN.
The $7\to 6$ at 4.6 keV and $6\to 5$ at 7.6 keV transitions were measured with
3 $\sigma$  and 3.7 $\sigma$  statistical significance, and the results
published in [2].
After this first measurement, a second one was performed in May-June 2002,
with an improved setup and better machine conditions (new optics which allowed
the increase of luminosity and the decrease of background).
The goals of this measurement were: to obtain a clean spectrum, in which
only kaonic nitrogen lines are present, to reduce the continuous background
with respect to the previous run (May 2001) in the view of the future kaonic
hydrogen measurement and to perform a fine-adjustment of the degrader, which
takes into account the $\phi$ production mechanism at a machine as DA$\Phi$NE
(i.e. the boost).
All these goals were successfuly reached.

The target used for this measurement, differently from before, was done in
kapton 75$\mu$m width, reinforced with carbon fiber. This had as positive
consequence the fact that in the X ray spectrum all the eventual electronic
transition lines corresponding to
``biasing signals", as iron, manganese, copper, were absent or much reduced, 
and the konic nitrogen signal was much clearer.
 The measurement was performed with the  target  filled with nitrogen
 at about 78 K and 0.98 bar ($\rho \simeq 4.3 \rho_{NTP}$). 
As detector,  16 CCD-55 were used.
A pure-background spectrum (so-called ``empty-target")
 was obtained as well, by using an over-dimensioned degrader,
stopping the kaons inside, so not allowing the formation of the kaonic nitrogen
atoms inside the nitrogen target. This spectrum was used for the shaping of the
 background and to obtain a background subtracted spectrum.

A preliminary analysis of the data gave as result an overall X ray kaonic
nitrogen spectrum which is shown in Figure 1, together with a fit in the region
of interest.
The empty-target subtracted spectrum is shown in Figure 2.
The calcium electronic transition is due to the presence of calcium in the
carbon fiber reinforcement of the target, while zirconium was placed, as a thin
foil, for calibration purpose.

The analysis (preliminary) of this spectrum gave a
 number of events in the 7.6 keV peak of $1400 \pm 132$ (10.6 $\sigma$
 statistical significance) and in the 4.6 keV peak of $700 \pm$142 
 (about 5 $\sigma$). More refined analysis are going to be performed.

The spectrum in Figure 1 shows the nice feature of a continuous backgrouns;
no electronic transitions are present in the region of interest (4-10 keV),
apart eventually  from some iron, which will be shielded in the next run.
Moreover, a reduction of background by a factor about 5 was reached with respect
to May 2001 run and the optimisation of the degrader was successfuly performed.
The kaonic nitrogen transitions were identified with a much better statistical
significance as well.

The next stage of the experiment is the measurement of the K-complex in 
kaonic hydrogen transitions.

\begin{figure}[htb]
\begin{center}
\mbox{\epsfig{file=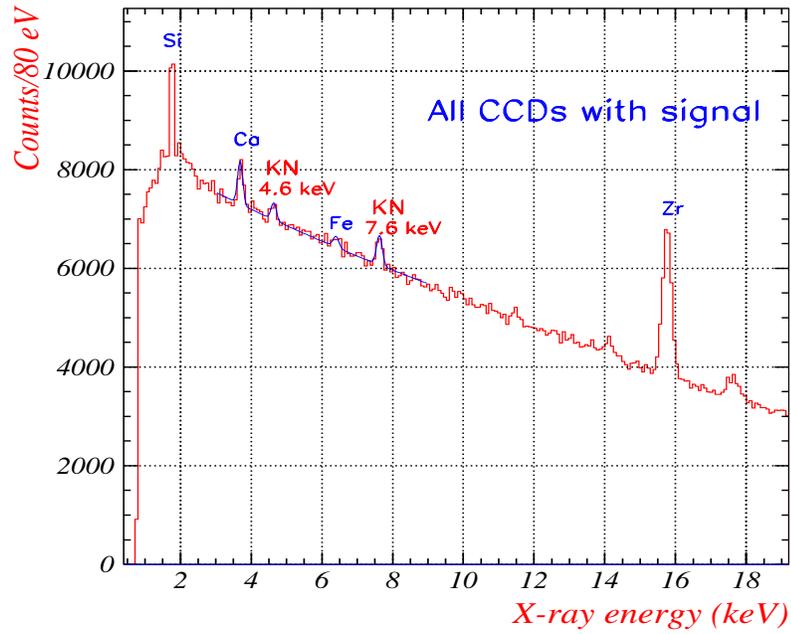,height=9.5cm,width=12cm}}
\caption{\small{The kaonic nitrogen X ray spectrum with a fit in the region of
interest.}}
\end{center}
\end{figure}

\begin{figure}[htb]
\begin{center}
\mbox{\epsfig{file=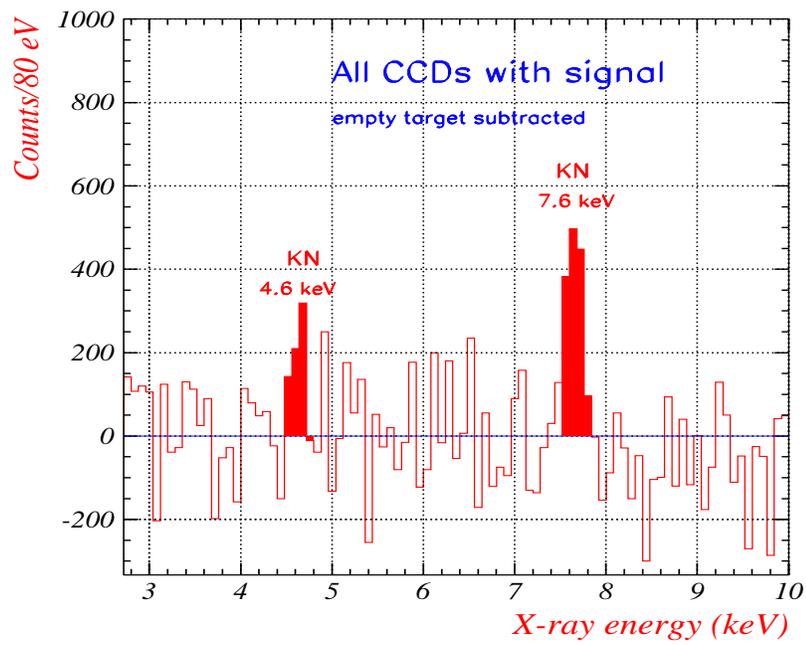,height=9.5cm,width=12cm}}
\caption{\small{The kaonic nitrogen empty-target  subtracted spectrum.}}
\end{center}
\end{figure}


\clearpage\label{gotta}\begin{center}
{\Large{\bf The Pionic-Hydrogen Experiment at PSI}}

\bigskip

{\bf   
D.\,F.\,Anagnostopoulos$^1$,
S.~Biri$^2$,
H.~Fuhrmann$^3$,
\underline{D.~Gotta$^4$},
M.~Giersch$^3$, 
A.~Gruber$^3$, 
A.~Hirtl$^3$,
M.~Hennebach$^4$,
P.~Indelicato$^5$,
Y.--W.~Liu$^6$,
B.~Manil$^5$,
V.~M.~Markushin$^6$,
N.~Nelms$^7$,
L.~M.~Simons$^6$,
P.~A.~Schmelzbach$^6$,
M.~Trassinelli$^5$,
and J.~Zmeskal$^3$
}\\[2mm]

$^1$~{\em Department of Material Science, University of Ioannina, GR--45110 Ioannina, 
          Greece,}
$^2$~{\em Institut of Nuclear Research; Hungarian Academy of Science, H-4001 Debrecen,
          Hungary}

$^3$~{\em IMEP, \"Osterreichische Akademie der Wissenschaften, A-1090 Vienna, Austria,}
$^4$~{\em Institut f\"ur Kernphysik, Forschungszentrum J\"ulich, D-52425 J\"ulich,}
$^5$~{\em Laboratoire Kastler--Brossel, Universit\'{e} Pierre et Marie Curie, 
          F--75252 Paris, France,}
$^6$~{\em Paul--Scherrer--Institut (PSI), CH--5232 Villigen, Switzerland,}
$^7$~{\em Department of Physics and Astronomy, University of Leicester, 
           Leicester LEI7RH, England}
\end{center}

In pionic hydrogen the hadronic pion--nucleon interaction manifests itself by 
a change of the energies and of the natural line width of X--ray lines
as compared to a purely electromagnetically bound atomic system. Experimentally 
accessible are the transitions to the 1s ground state emitted in
the last de--excitation step of the atomic cascade. Any observed 
strong--interaction effect can be attributed fully to the 1s state, because 
2p--state effects are negligibly small. Hence, the s--level shift and width 
are exclusively owing to the pion-nucleon s--wave interaction, which is 
described in the limit of isospin symmetry by the isoscalar and isovector 
scattering lengths $a^{+}$ and $a^{-}$ [1].

To improve on the accuracy for the hadronic parameters as compared to previous 
measurements [2], a thorough study of a possible influence of de--excitation 
processes is essential. A first series of measurements has been completed 
by the new pionic--hydrogen experiment at the Paul--Scherrer--Institut (PSI), 
using the new cyclotron trap, a cryogenic target and a Bragg crystal spectrometer 
equipped with spherically bent silicon and quartz crystals and a large--area 
CCD array [3]. Data analysis is in progress.

In order to identify radiative de--excitation of the $\pi H$ system -- when 
bound into complex molecules formed during collisions 
$\pi^{-} p+H_{2}\rightarrow [(pp\pi^{-})p]ee$ [4] -- the
energy of the $\pi H(3p-1s)$ transition was measured at various 
target densities. X--ray transitions from molecular states should show up as 
low--energy satellites with density dependent intensities because of 
different collision probabilities. In our experiment, 
covering the pressure range from 3.5 bar to liquid, no density effect could 
be established [5]. It is concluded that the decay of molecules is dominated
by Auger emission. The value derived for the hadronic shift is in
agreement with the result of the previous experiment. 

At present, the accuracy for the hadronic broadening (7\%) is limited by a 
not precisely known correction to the measured line width originating from the
Doppler broadening due to Coulomb de--excitation [2]. For that reason the 
precisely measured 1s--level shift in pionic deuterium was used together 
with the shift of hydrogen in the determination of the $\pi N$ scattering 
length [6]. This procedure, however, requires a sophisticated treatment 
of the 3--body system $\pi D$. In addition, up to now it cannot be excluded
that the radiative decay channel after molecule formation is enhanced in 
deuterium compared to hydrogen.

To improve the data base on the line broadening due to Coulomb de--excitation, 
the three $\pi H(2p-1s)$ (2.4~keV), $\pi H(3p-1s)$ (2.9~keV) and $\pi H(4p-1s)$ 
(3.0~keV) transitions were studied at a target density equivalent to 10~bar. 
An increase of the line width was found for the $2p-1s$ line compared to the 
$3p-1s$ transition, which is attributed to the higher energy release available 
for the acceleration of the pionic--hydrogen system. This result is corroborated 
by a reduced line width of the $4p-1s$ line. For the total line width 
of the $\pi H(3p-1s)$ transition an increase of 5-10\% was found compared to 
the result of Schr\"oder et al. [2], which may be due to the significantly 
improved background conditions in the new experiment.

The response of the crystal spectrometer was obtained from the
$\pi^{12}C(5g-4f)$ line (3.0~keV), which is negligibly narrow compared to
the experimental resolution (Fig.~1). The statistics of such a measurement, 
however, and with that the accuracy for the response function, is limited 
due to beam time considerations. In addition, the crystal response has been
measured with X--rays emitted from helium--like Ar, ionised by means of an 
electron--cyclotron resonance ion trap (ECRIT) set up at PSI [7]. With that,
studies of the Bragg crystals used up to now with good statistics could be 
performed within a reasonable time scale.

\begin{figure}[h]
\begin{center}
\includegraphics[width=15cm]{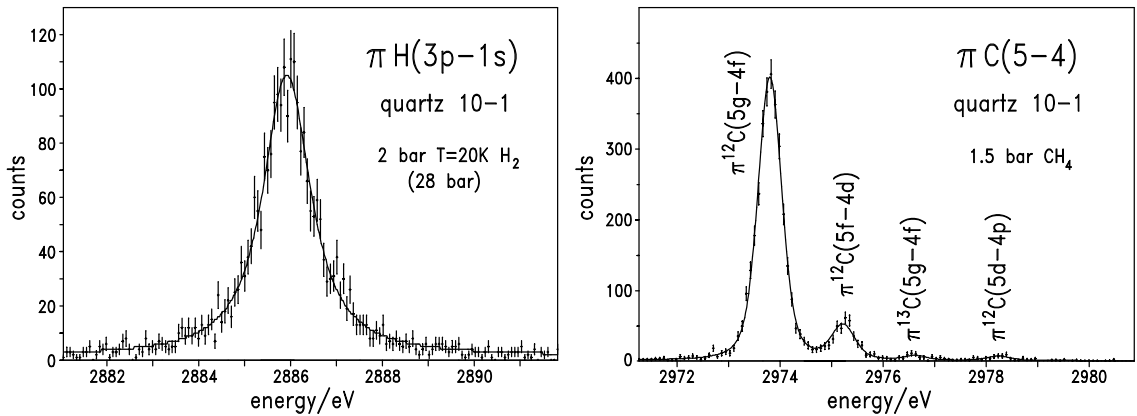}
\caption{Fig.~1: Ground--state transition $3p-1s$ in pionic hydrogen and the 
         pionic carbon $5-4$ transitions measured with a quartz crystal.}
\end{center}
\end{figure}

>From about 2005 on, Coulomb de--excitation will be studied in detail in the 
absence of strong--interaction effects by measuring K transitions from muonic 
hydrogen. Together with the detailed knowledge of the response function 
by using the ECRIT and a newly developed cascade code [8], which includes
the velocity dependence of the atomic cascade, a sufficiently accurate 
correction for the Doppler broadening in pionic hydrogen should be
achieved to extract the hadronic broadening at the level of about 1\%.


\clearpage\label{yazkov}\begin{center}
{\Large{\bf Lifetime measurement of $\pi^+\pi^-$ atom at DIRAC}}
\bigskip

{\bf V.~Yazkov} on behalf of DIRAC Collaboration\\[2mm]

{\em Skobeltsyn Institute for Nuclear Physics of Moscow State University}

\end{center}

\subsubsection*{Introduction}

Pionium or $A_{2\pi}$ is a hydrogen-like atom consisting of $\pi^+$ and 
$\pi^-$ mesons. The lifetime of this atom is inversely proportional to the 
squared difference between the S-wave $\pi\pi$ scattering lengths for isospin 
0 and 2, $|a_0 -a_2|$. This value is predicted by chiral perturbation theory 
(ChPT) [1], and a measurement of the $\pi^+\pi^-$ atom lifetime provides 
a possibility to check predictions of ChPT in a model-independent way. 

\subsubsection*{Method of lifetime measurement}

The $A_{2\pi}$ are produced by Coulomb interaction in the final state 
of $\pi^+\pi^-$ pairs generated in proton--target interactions [2,3]. 
After production $A_{2\pi}$ travel through the target and 
some of them are broken up due to their interaction with matter: ``Atomic 
pairs'' are produced, characterised by small pair c.m. relative momenta 
$Q < 3$~MeV/$c$. These pairs are detected in the DIRAC setup. Other atoms 
annihilate into $\pi^0\pi^0$. The amount of broken up atoms $n_A$ depends 
on the lifetime which defines the decay rate. Therefore, the breakup 
probability is a function of the $A_{2\pi}$ lifetime.
The dependence of $P_{\rm br}$ on the lifetime $\tau$ is determined by 
the solution of differential transport equations [4].

Also $\pi^+\pi^-$ pairs are generated in free state. Essential fraction of 
such pairs (``Coulomb pairs'') are affected by Coulomb interaction, too. 

The aim of DIRAC is to measure the 
$A_{2\pi}$ breakup probability $P_{\rm br}(\tau)$. $P_{\rm br}(\tau)$ is 
the ratio between the observed number of ``atomic pairs'' and the number of 
produced $\pi^+\pi^-$ atoms which is calculated from the measured number of 
``Coulomb pairs''.

\subsubsection*{Experimental setup}

The purpose of the DIRAC setup [5] is to detect $\pi^+\pi^-$ 
pairs with small relative momenta. This setup is 
located at the CERN T8 beam area (East Hall). It became operational at 
the end of 1998 and uses the 24~GeV proton beam from PS accelerator. 

The setup resolution over the relative c.m. momentum $Q$ of $\pi^+\pi^-$ pair
is better than 1~MeV/$c$.



\subsubsection*{Experimental data}

\begin{figure}[htb]  
\begin{center}
\epsfxsize=11.0cm\epsfysize=12.0cm\epsfbox{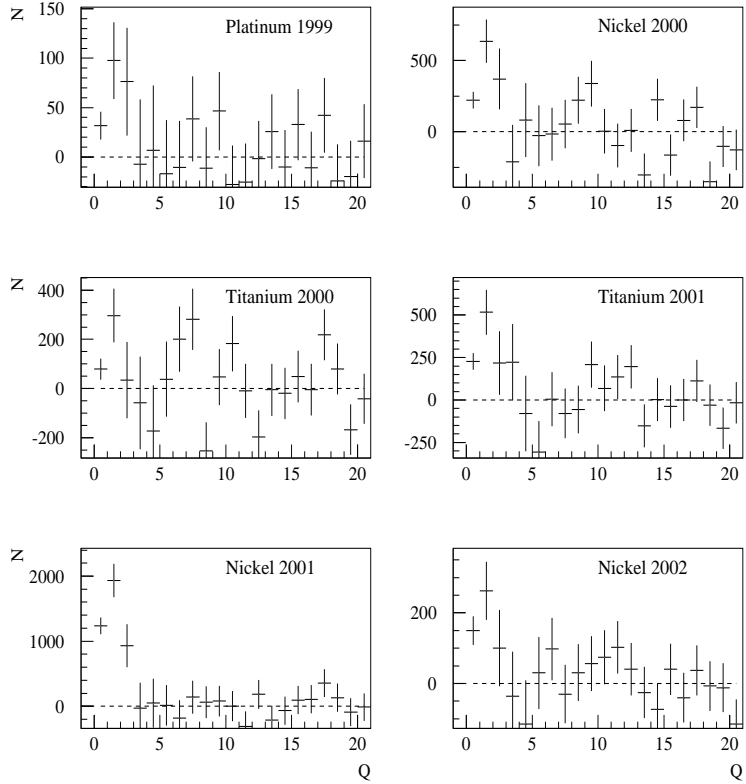}
\end{center}
\vspace*{-4.ex}
\caption{Signals of ``atomic pairs'' in the region $Q<3$~MeV/$c$ 
are obtained with different targets. For 2002 only data collected 
in 15 days are analysed.}\label{feff}
\end{figure}

The experimental $Q$-distribution of $\pi^+\pi^-$ pairs is fit by 
an approximated distribution of ``free pairs'' in the region 
$Q>3.5$~MeV/$c$ where ``atomic pairs'' are absent. 
The number of ``atomic pairs'' is obtained as excess of experimental 
$Q$-distribution above the approximated distribution of ``free pairs'' 
only in the region $Q<3.5$~MeV/$c$. 
Differences of experimental and approximated distributions are shown in
Fig.~\ref{feff} for data collected in 1999, 2000, 2001 and in 15 days of
2002. Platinum,  nickel and titanium targets havebeen used.

The overall statistics of ``atomic pairs'' which is collected before 
the end of 2001 is more than 5000. It provides statistical 
accuracy for the lifetime measurement at the 20\% level. 

The data analysis procedure is not completed yet and only very preliminary 
lifetime estimations are $\tau=2.8^{+1.1}_{-0.8} \cdot 10^{-15}$s 
for data collected in 2000 with nickel target and 
$\tau=5.4^{+1.5}_{-1.3} \cdot 10^{-15}$s for all titanium target data. 
Averaged value is $\tau=3.6^{+0.9}_{-0.7} \cdot 10^{-15}$s.

\subsubsection*{Conclusions}

Preliminary results have been achieved by analysing a sample of 5000 atoms. 
The statistical accuracy in the lifetime determination reaches 20\%.
Data collected in 2002 allow us to increase the statistical accuracy up to
15\%.


\clearpage\label{cheshkov}\begin{center}
{\Large{\bf Prospects for the study of the $K^\pm_{e4}$ decays at the NA48/2 experiment}}

\bigskip

{\bf \underline{C.~Cheshkov} and G.~Marel}\\[2mm]  

{\em DSM/DAPNIA - CEA Saclay, F-91191 Gif-sur-Yvette, France}

\end{center}

It is well known that the study of the $K^\pm\to\pi^+\pi^-e^\pm\nu_e$
decay parameters can provide an important information about the
$\pi\pi$ interaction near threshold. Besides the extraction of the
decay form-factors which are valuable input for $\chi$PT, the
measurement of the $\pi\pi$ scattering length $a^0_0$ is recognized as
a crucial cross-check with the current understanding of chiral symmetry
breaking of QCD and as an accurate estimate of the size of quark
condensate $<0|u\overline{u}|0>$ [1].
The two most significant experiments so far were carried out by the
Geneva-Saclay [2] and E865 collaborations [3] and are based on $3\times
10^4$ and $4\times 10^5$ collected $K^\pm\to\pi^+\pi^-e^\pm\nu_e$ events.
Since the present experimental uncertainty on $a^0_0$ is significantly
larger than the theoretical one it is of great interest to acquire an
additional high statistics and good quality experimental data in this
decay mode.

The NA48/2 experiment is approved as an extension of the NA48
experiment at the CERN SPS.
Its main goals are the search for direct CP
violation in the Dalitz plot asymmetries in the decays of opposite
charged kaons, the measurement of the scattering length $a^0_0$ and the
analysis of various rare kaon decays. About $1\times 10^{11}$ $K^+$ and
$0.6\times 10^{11}$ $K^-$ decays are expected for around 100 days of
data taking in 2003.
The NA48/2 experiment is based on a slightly modified NA48
set-up. The kaon decays volume is contained in a 114m long vacuum tank
which is followed by the NA48 detectors. The
charged particles are detected and reconstructed by a high resolution
magnetic spectrometer. It consists of four drift chambers and a dipole
magnet with horizontal momentum kick of 120MeV/c. The corresponding
track momentum resolution is given by $\sigma_P/P [\%]\approx
0.48\oplus 0.02\times P[GeV/c]$. The energy, position and time of the
showers produced by the charged particles passed through the
spectrometer are measured by a quasi-homogeneous liquid krypton
electromagnetic calorimeter (LKr). Its transverse structure of
$\approx$1300 readout cells each with size of $2\times 2$cm$^2$ is formed by
electrodes extended with an accordion geometry. LKr
has a longitudinal size of 127cm (equivalent to $\approx
27 X_0$ and $\approx 2\lambda$) and a projective tower geometry
pointing to the decay volume. Its energy resolution for
e/m showers is $\sigma_E/E=0.42\%\oplus 3.2\%
/\sqrt{E}\oplus 0.09/E$ ($E$ in GeV). A detailed description of the
whole NA48 set-up can be found elsewhere [4].

In addition to the present NA48
apparatus, two essential new elements will be introduced to the
NA48/2 set-up [5]. The first one is an achromatic beam-line capable to
transport and focus simultaneously 60GeV/c$\pm$10$\%$ $K^+$ and $K^-$
beams. A kaon beam spectrometer (KABES) will measure the
kaon beam sign, momentum and direction in order to improve the
reconstruction of the $K^\pm_{e4}$ and to recover the 
$K^\pm\to 3\pi$ with one pion escaping
detection [5]. KABES consists of three so-called
stations - two in the beam-line achromat and one downstream. Each
station is formed by two time projection MICROMEGAS chambers [6]
with opposite drift directions. The detector prototype has been
successfully tested in July 2002 in the environment of an extremely high
particle rate of 20-30MHz (comparable to that of the future kaon
beam-line) showing excellent time ($<$1ns) and space ($<$80$\mu$m)
resolutions.

Because of the small branching ratio of $K^\pm\to\pi^+\pi^-e^\pm\nu_e$
an important point in the systematic free measurement of $a^0_0$ is
the reduction of the major background coming from
$K^\pm\to\pi^+\pi^-\pi^\pm$ decays with $\pi^\pm$ misidentified as an
electron. The situation is even more complicated since the background
populates the low $\pi\pi$ invariant mass region which is the most 
sensitive to $a^0_0$.
To cope with the relatively large expected $K^\pm\to\pi^+\pi^-\pi^\pm$
background the NA48/2 collaboration has introduced a {\it Neural Network}
(NN) based approach [7] of exploiting the whole available LKr
information about the longitudinal and transverse electromagnetic and
hadronic showers development. Each electron candidate is described by
the charged track
measured by the spectrometer and the associated LKr shower parameters
and is passed then as an input to a specially designed multilayer
NN. The candidate is identified as an electron or rejected as a pion
taking into account the probability of electron hypothesis given by
NN output. The NN was developed,
trained and tested using pure experimental samples of electrons and
charged pions. By applying the described above method a pion
rejection factor of $3500$ at $94\%$ electron detection efficiency is achieved.

In order to estimate the expected acceptance, resolution and
background for $K^\pm_{e4}$ decays a GEANT based Monte-Carlo simulation
of the apparatus has been used. The $K^\pm\to\pi^+\pi^-e^\pm\nu_e$
decay has been described by three real form-factors
($\overline{g}=g/f_s$, $\overline{g}^\prime=g^\prime/f_s$,
$\overline{h}=h/f_s$) and one phase ($\delta=\delta_0^0 -
\delta_1^1$). The values for these parameters were taken from [2]. The
scattering length $a^0_0$ has been extracted following the procedure
in [2]. The selection of $K^\pm\to\pi^+\pi^-e^\pm\nu_e$ events has been
made by applying a set of standard acceptance and kinematic cuts
(including a cut on the transverse momentum of the three charged
tracks) as well as the NN electron identification. The expected
acceptance of $\approx 30\%$ has been found
relatively flat upon the Cabibbo-Maksymowicz variables leading to a conclusion
of small systematic effects due to the inexact knowledge of the
detector response. Taking into account the expected charged kaon flux
in 2003 the expected $K^\pm\to\pi^+\pi^-e^\pm\nu_e$ decay statistics
is $>1\times 10^6$ and the corresponding statistical uncertainty on
$a^0_0$ is less than 0.01. 
The estimated resolution and background contaminations are presented
in Table \ref{tab1} and Table \ref{tab2}, respectively.
\begin{table}[t]
\begin{tabular}{c|c|c|c}
RMS & Geneva- & E865 & NA48/2 \\
    & Saclay  &      & (expected) \\
\hline
$M_{\pi\pi}$ (MeV) & 2.5 & 16 & 1.6 \\
$M_{e\nu}$ (MeV) & 6.3 & 25 & 5.1 \\
$\theta_{\pi\pi}$ (mrad) & 50 & 63 & 37 \\
$\theta_{e\nu}$ (mrad)   & 40 & 47 & 47 \\
$\phi$ (mrad) & 157 & 172 & 168 \\
\end{tabular}
\hfill
\begin{tabular}{c|c}
Mode & Background \\
\hline
$K^\pm\to\pi^+\pi^-\pi^\pm$ ($\tau$) & $\le 1\%$ \\
$K^\pm\to\pi^+\pi^-\pi^\pm\to e^\pm\nu$ & $<0.4\%$ \\
$K^\pm\to\pi^\pm\pi^0\to e^+e^-\gamma$  & $<0.3\%$ \\
$K^\pm\to\pi^\pm\pi^0\pi^0\to e^+e^-\gamma$ & $<0.1\%$ \\
$K^\pm\to e^\pm\nu\pi^0\to e^+e^-\gamma$ & $<0.1\%$ \\
$K^\pm\to\mu^\pm\nu\pi^0\to e^+e^-\gamma$ & $<0.1\%$ \\
$K^\pm\to e^\pm\nu\pi^+\pi^-\to\mu\nu$ & $<0.2\%$ \\
\end{tabular}
\parbox[t]{0.52\textwidth}{\caption{Resolution on Cabibbo-Maksymowicz variables.}\label{tab1}}
\hfill
\parbox[t]{0.47\textwidth}{\caption{Background contaminations.}\label{tab2}}
\end{table}
As one can
see the extremely good performance of the charged particles
spectrometer reflects in the high resolution on invariant mass
variables. On the other hand, the relatively high beam momentum of
60GeV/c leads to moderate resolutions upon the angle variables which
are nevertheless comparable to those in the previous low energy
experiments. Due to the high $e/\pi$ separation provided by the NN
method the $K^\pm\to\pi^+\pi^-\pi^\pm$ background is expected to be
less than 1$\%$. A detailed analysis of this background has shown that
the corresponding systematic effect in the measurement of $a^0_0$ is
less then 0.01.

\begin{figure}[t]
\begin{center}
\includegraphics[width=5cm]{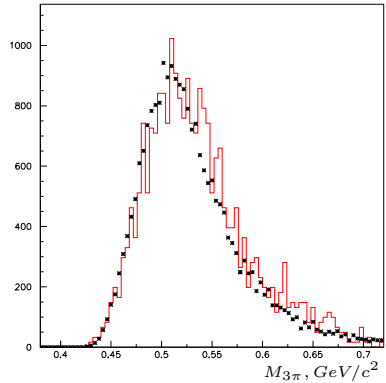}
\hspace{-1.9cm}
{\tiny $M_{3\pi}, GeV/c^2$}
\hspace{1.9cm}
\caption{The $3\pi$ invariant
mass distributions for the data (solid line) and MC
(stars).}\label{fig2}
\end{center}
\end{figure}

In the framework of the preparation for the future experiment, in September
2001 the NA48/2 collaboration has performed a 4 hours charged kaons
beam test run. The already tuned $K^\pm_{e4}$ decays selection and
reconstruction has been applied to the collected during this run
data. 1477 events passed all the selection criteria including the NN
electron identification procedure. The corresponding branching ratio
has been evaluated using $K^\pm\to\pi^+\pi^-\pi^\pm$ decay as a
normalization channel. The obtained preliminary result of BR$=(3.75\pm
0.25)\times 10^{-5}$ is in good agreement with the world average
value [8]. A careful study of the differences between the data and
Monte-Carlo $M_{3\pi}$ distributions has confirmed that the $K^\pm\to 3\pi$
background is less than $1\%$ (Figure \ref{fig2}). 

To summarize, the NA48/2 experiment is planning to collect $>1\times
10^6$ $K^\pm\to\pi^+\pi^-e^\pm\nu_e$ decays for one year data taking
in 2003 and to measure the $\pi\pi$ scattering length $a^0_0$ with a
statistical precision of $<$0.01. The excellent $e/\pi$
separation power combined with the good detector resolution will allow
to keep and to control the systematic uncertainty to the level of less
than 0.01. Thus the future measurement of $a^0_0$ by the NA48/2
together with the forthcoming results from the DIRAC experiment [9]
are expected to improve significantly the present experimental picture
on this important parameter.


\clearpage\label{nemenov}
\begin{center}
  {\Large\bf Future experimental investigation of the
    \boldmath{$\pi^+\pi^-$} atom}

\bigskip

{\bf L. Nemenov}\\[2mm]  

{\em CERN, CH-1211, Geneva 23, Switzerland and JINR, Dubna, Russia}
\end{center}

The difference $\Delta E_n$ between the energy of atomic $ns$ and $np$
states of $A_{2\pi}$ includes the contributions of the vacuum
polarisation $\Delta E_n^{vac}$, and of the strong interaction $\Delta
E_n^{s}$ effects, where $\Delta E_n^{s} \sim 2a_0+a_2$, and $a_0$ and
$a_2$ are the $s$-wave $\pi\pi$ scattering lengths with isotopic-spin
quantum numbers 0,2 [1,2].  The value of $\Delta E_n^{vac}$ is well
known from QED calculations [1,3].  For this reason the measurement of
$\Delta E_n$ will give, in a model independent way, the value of
$2a_0+a_2$ [4,5].

Some possibilities are discussed for measuring the $np-ns$ energy
splitting in ($\pi^+\pi^-$) atoms in experiments with relativistic
particle beams, using the electromagnetic field control of
annihilation process [6,7]. Significant resonant enhancement of
annihilation probability may be observed in an oscillating field. The
resonance provides much more spectacular changes in the decay rate
than corresponding decay rate changes in a steady field. The position
of resonance on the frequency axis gives an additional and more
accurate information on the np-ns transition energy, than the data on
the ratio between the field-free and dc-field-induced annihilation
rates. Numerical results are presented for the states $n=2$ and $n=3$.


\clearpage\label{sazdjian}\begin{center}
{\Large{\bf Relevance of $K\pi$ atom measurements}}

\bigskip

{\bf H. Sazdjian}\\[2mm]  

{\em Groupe de Physique Th\'eorique, Institut de Physique Nucl\'eaire,\\
Universit\'e Paris XI, F-91406 Orsay Cedex, France}
\end{center}

It is pointed out that measurements of the lifetime and of the
$2s-2p$ energy level splitting may provide information about the
combinations $(a_0^{1/2}-a_0^{3/2})$ and $(2a_0^{1/2}+a_0^{3/2})/3$
of the strong interaction $S$-wave scattering lengths, with isospins
1/2 and 3/2, respectively. Corrections to the nonrelativistic formulas of
these quantities can be estimated with the use of the chiral effective
lagrangian in the presence of electromagnetism. The combination 
$(a_0^{1/2}-a_0^{3/2})$ is not sensitive at leading order to the chiral
symmetry breaking mechanism, but may provide information about the low
energy constant $L_5$. The combination $(2a_0^{1/2}+a_0^{3/2})/3$ is
sensitive to the quark condensate value and to Zweig rule violating 
effects through its dependences upon the low energy constants 
$(2L_6+L_8)$ and $L_4$. The electromagnetic and isospin breaking 
corrections are estimated to be of the order of a few percent in the 
lifetime value. For the $2s-2p$ energy level splitting, the first main 
correction comes from vacuum polarization with a 25\% effect. The next 
corrections come from the other electromagnetic and isospin breaking 
effects and are of the order of a few percent.

\clearpage\label{buettiker}\begin{center}
{\Large{\bf The Roy equations for the ${\mathbf\pi K}$ system}}

\bigskip

{\bf P.~B\"uttiker$^1$, S.~Descotes$^2$, and B.~Moussallam$^{1}$}\\[2mm]  

$^1$ {\em Institut de Physique Nucl{\'e}aire, Universit{\'e} de
  Paris--Sud, F-91406 Orsay} 

$^2$ {\em Laboratoire de Physique Th{\'e}orique Hautes Energies, Universit{\'e} de
  Paris--Sud,\\F-91406 Orsay}

\end{center}
In the past, $\pi K$ scattering attracted the
interest of experimentalists as well as theorists. Unfortunately, the
theoretical analyses of this process often suffered from the low
statistics of the experiments and sometimes from over-simplified
theoretical assumptions. On the other hand high statistics
experimental data became available in the late seventies and eighties
which have never been subject of a stringent theoretical analysis.
Recently, there has been a revival in the interest in $\pi K$
scattering: there are indications for a flavour dependence of the size
of the quark condensate [1], and as the $\pi K$ system is the most
simple $SU(3)$--process involving kaons, this process is a suitable
place to test this dependence; effects of isospin violation in the
above process are studied in the framework of ChPT [2] in order to obtain
reliable predictions for the two $S$--wave scattering lengths from the
$\pi K$ bound state experiment of DIRAC [3]; the $\pi K$
$\sigma$--term has been analyzed [4] which can be regarded as an
intermediate step towards ambitious calculations for the $K N$ system;
and finally $\pi K$ scattering data were used to investigate the
existence of a $\kappa$ resonance [5].

However, all these calculations strongly depend on reliable
experimental data at very low energies. Unfortunately, the available
high statistics data start only at about $800$ MeV so that one is left
with a gap at low energies. It has been shown for the $\pi\pi$ system
that solving the Roy-equations is a suitable tool to close this gap [6].
Here we apply this method to the $\pi K$ system.

Assuming isospin to be a conserved quantity, $\pi K$ scattering can be
described in terms of the isospin--even and --odd amplitudes
$F^+(s,t)$ and $F^-(s,t)$, respectively, the crossed channel,
$\pi\pi\to K\bar{K}$ is given by $F^{I_t=0}(s,t)$ and
$F^{I_t=1}(s,t)$. By combining fixed--$t$ and hyperbolic dispersion
relations with at most two subtractions for each of these amplitudes,
one can derive twice--subtracted dispersion relations where the two
$S$--wave scattering lengths $a^+_0$ and $a^-_0$ play the role of the
unknown subtraction constants. Projecting these dispersion relations onto
partial waves, together with unitarity, yields a set of coupled
integral equations for the $s$-- and $t$--channel partial waves
(Roy--equations) with two a priori free parameters $a^+_0$ and $a^-_0$
[7], e.g.
\begin{eqnarray*}
      g^1_1(t) & = &\frac{2\sqrt{2} m_+ a^-_0}{3(m_+^2-m_-^2)} +
                  \frac{t}{\pi}\int_{4 m_\pi^2}^\infty
                  \frac{d\,t'}{t'}
                  \frac{{\rm Im\,} {g^1_1(t')}}{t'-t}\\
               &  &  +\frac{1}{\pi}\int_{m_+^2}^\infty
                  d\,s'\left\{G^-_{1 0}(t,s'){\rm Im\,}f^-_0(s') +
                  G^-_{11}(t,s'){\rm Im\,}f^-_1(s')\right\} 
                  + d^1_1(t),
\end{eqnarray*}
where the $G^-_{ij}$ are the kernel functions and $d^1_1$ is a
so--called driving term accounting for the contributions of the higher
partial waves.\\
Using the available experimental data for the $s$-- and $t$--channel
above $1$ GeV and the driving terms as input, one can show --- in
analogy to the $\pi\pi$--system, see [6] and references therein ---
that the system of Roy--equations has a unique solution below $1$ GeV.

The equations then can be solved in several steps: using a
Schenk--type parametrization for the $s$--channel waves and due to extended
unitarity the $t$--channel partial waves can be solved by applying
Omn{\`e}s--Muskhelishvili methods. These solutions for the
$t$--channel waves then are used in $s$--channel equations and by
tuning the parameters of the $s$--channel waves one tries to find the
best input/output agreement for the Roy--equations. This procedure
generates approximative solutions of the $\pi K$ Roy--equations for
some ranges of values for $a^+_0$ and $a^-_0$. However, most of the
solutions are not physical and preliminary results indicate that in
$\pi K$ scattering the Roy--equations do not support a universal band,
which is a relation between the two $S$--wave scattering lengths, but
one definite point in the $a^+_0$--$a^-_0$--plane. Furthermore, these
results indicate a value for $a^-_0\equiv (a^{1/2}_0 - a^{3/2}_0)/3$
which is somewhat larger than what has been predicted by
$SU(3)$--ChPT [8], calling thereby for an independent experimental
check of the value of $a^-_0$ by the $\pi K$ bound state experiment of
the DIRAC collaboration [3]. 


\clearpage\label{descotes}\begin{center}
{\Large{\bf $\pi\pi$ scattering and\\
the chiral structures of QCD vacuum}}

\bigskip

{\bf \underline{S.~Descotes-Genon}$^{1,2}$, N.~Fuchs$^3$, 
L.Girlanda$^4$ and J.~Stern$^5$}\\[2mm]  

$^1$ {\em Laboratoire de Physique Th\'eorique, 91405 Orsay Cedex, France}

$^2$ {\em University of Southampton, SO17 1BJ Southampton, UK}

$^3$ {\em Purdue University, West Lafayette IN 47907, USA}

$^4$ {\em European Center for Theoretical Studies in Nuclear Physics,
38050 Trento, Italy}

$^5$ {\em Institut de Physique Nucl\'eaire, 91405 Orsay Cedex, France}
\end{center}


At low energies, QCD is dominated by the spontaneous
breakdown of chiral symmetry which leads to the appearance of
pseudoscalar Goldstone bosons $\pi,K,\eta$.
Due to the mass hierarchy $m_{u,d}\ll m_s\ll \Lambda_{H}$,
we can consider two different chiral limits:
\begin{equation}
N_f=2:\, m_u,m_d\to 0\,,\ m_s\ {\rm physical}\,,\qquad
N_f=3:\, m_u,m_d,m_s\to 0\,.
\end{equation}
In both limits, chiral order parameters can be defined, such as
the quark condensate: $\Sigma(N_f)=\lim_{N_f}\ \langle 0|\bar{u}u|0\rangle$,
and the pseudoscalar decay constant: 
$F(N_f)=\lim_{N_f} F_\pi$.

The patterns of chiral symmetry breaking reflected by these
parameters can be quite different in the $N_f=2$ and $N_f=3$
limits~[1]. For instance, the condensates are related through:
$\Sigma(2)=\Sigma(3)+ m_s Z_{\rm scalar}$,
where $Z_{\rm scalar}>0$ describes the violation
of the Zweig rule in the scalar sector $0^{++}$. 
There have been recently
hints of a significant decrease from $\Sigma(2)$ to $\Sigma(3)$,
which would lead to a different low-energy behaviour of the $N_f=2$ ($\pi\pi$
scattering) and $N_f=3$ ($\pi K$ scattering) sectors.

Because of the possibility of a small three-flavour condensate $\Sigma(3)$, 
more care is needed when dealing with chiral expansions obtained
from $N_f=3$ chiral perturbation theory, e.g.:
\begin{eqnarray} \label{eq:mpi}
F_\pi^2 M_\pi^2 &=& 2m\Sigma(3) +  
     m_q^2 [b^q_{\pi Q}\log(M_Q/\mu) + c^q_\pi(\mu)] 
  + F_\pi^2M_\pi^2 d_\pi\,,\\
F_K^2 M_K^2 &=& (m+m_s)\Sigma(3) +  
     m_q^2 [b^q_{K Q}\log (M_Q/\mu) + c^q_K(\mu)]
  + F_K^2M_K^2 d_K\,. \label{eq:mka}
\end{eqnarray}
Next-to-leading-order (NLO) terms contain chiral logarithms ($Q=\pi,K,\eta$),
and two low-energy constants $L_8$ and $L_6$ (in $c_P$).
$L_6$ is related to $Z_{\rm scalar}$ and
measures how badly the Zweig rule is violated in the scalar sector.
The remainders $d_P$ collect NNLO and higher-order terms.

We want to take into account the possibility of a suppressed
three-flavour condensate $\Sigma(3)$. We will assume 
an overall convergence of the chiral series [NNLO remainders
$d_\pi,d_K\sim (30\%)^2 \sim 10\%$], but we do \emph{not}
make assumptions about the relative sizes of leading- and 
next-to-leading-order terms. We simply consider the chiral 
series (\ref{eq:mpi}) and (\ref{eq:mka}) as exact identities.

We can first express the 
three-flavour quark condensate in terms of $F(3)$, $r$
and $L_6$~[1]:
\begin{equation} \label{eq:x3}
X(3) = \frac{2m\Sigma(3)}{F_\pi^2M_\pi^2}
  = 2\frac{1-\epsilon(r)-d}{1+\sqrt{1+\xi}}\,, \qquad 
\xi = 64\frac{M_\pi^2}{F_\pi^2}(r+2)
   \frac{1-\epsilon(r)-d}{Z(3)} (L_6-L_6^{\rm crit})\,,
\end{equation}
where $r=m_s/m$, $Z(3)=[F(3)/F_\pi]^2$,
$\epsilon(r)<0.2$ for $r>15$ and
$L_6^{\rm crit}$ depends (weakly) on r only
($L_6^{\rm crit}(M_\rho)=-0.26\cdot 10^{-3}$ for $r=25$).
The NNLO remainder $d$ is estimated as $d\sim d_\pi\sim 10\%$.
The square root arising in eq.~(\ref{eq:x3}) indicates the nonperturbative
resummation of Zweig-rule violating effects. It is
expanded in the standard treatment of $\chi$PT, assuming
$\xi\ll 1$ -- inspired by large-$N_c$ expectations. But a small
shift of $L_6$ from $L_6^{\rm crit}$ of a few $10^{-3}$ would
be magnified by the huge scaling factor 
$64 M_\pi^2(r+2)/F_\pi^2\sim 4\cdot 10^3$, leading to a damping
of $X(3)$.

For the two-flavour quark condensate, the chiral expansions
of $F_\pi^2M_\pi^2$ and $F_K^2M_K^2$ 
yield~[2]:
\begin{equation}
X(2)[1-\bar{d}_\pi]=X(3)+
  \frac{r}{r+2}
      \Big\{[1-X(3)]-\epsilon(r)-d-f[X(3)/Z(3)]^2
   \Big\}\,,
\end{equation}
where $\bar{d}_\pi\sim d_\pi\pm 3\%$ and
$f$ is a function of chiral logarithms in the $N_f=2$ chiral limit
($0.03\leq f \leq 0.06$). The two-flavour quark condensate
$X(2)$ is therefore the sum of two terms: a genuine condensate
$X(3)$ and an induced condensate proportional to $m_s$.
For large $r$ ($r>15$), $X(2)$ remains close to 1 for any size of 
$X(3)$ due to 
a compensation from the induced condensate. This phenomenon can 
be interpreted in terms of the density and fluctuation of small
eigenvalues of the Euclidean Dirac operator~[2].

$X(2)$ is weakly correlated with the 
pattern of $N_f=3$ chiral symmetry breaking, but strongly correlated
with the value of the quark mass ratio $r$~[1], as seen
from the plot on the left.

\begin{center}
\includegraphics[width=6cm,angle=270]{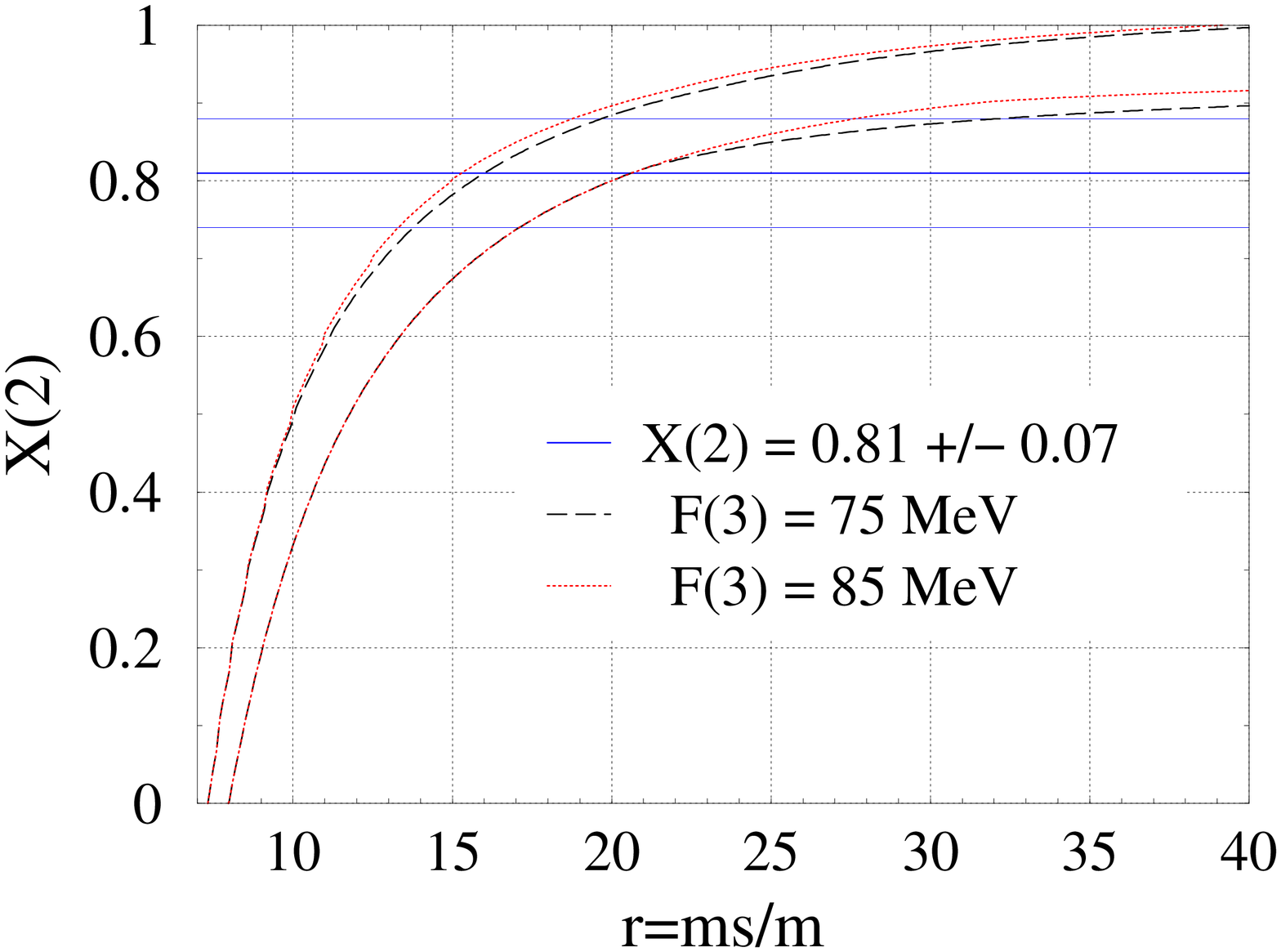}\quad
\includegraphics[width=6cm,angle=270]{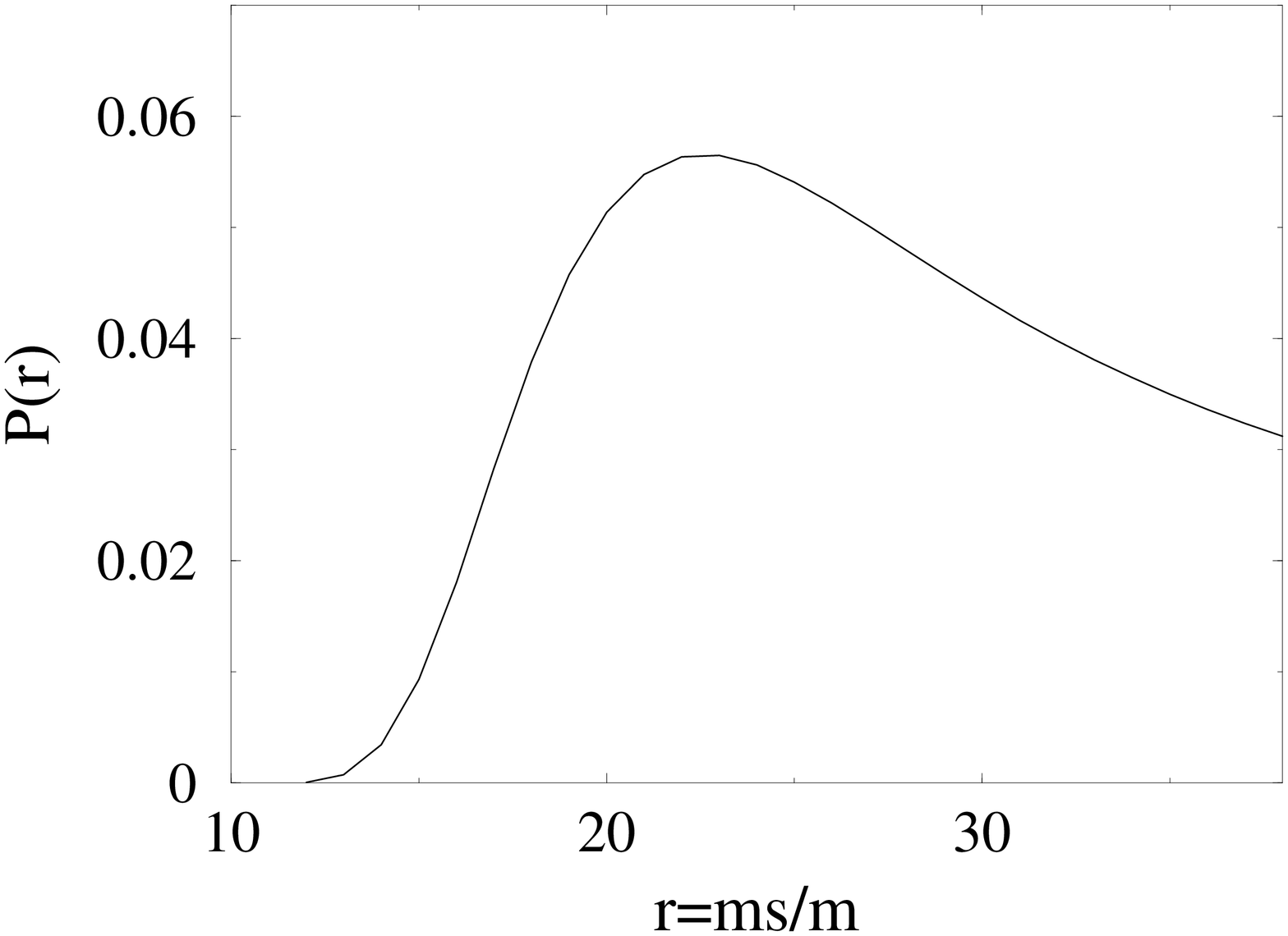}
\end{center}

Accurate measurements of $\pi\pi$ scattering would pin down
$N_f=2$ chiral order parameters, but would only constrain
mildly the pattern of $N_f=3$ chiral symmetry breaking.
The analysis of available $\pi\pi$ data~[3] leads to $X(2)=0.81\pm 0.07$, 
implying that $r\geq 15$. The Bayesian approach allows one to express more
quantitatively this statement and to combine results
concerning different observables~[4]. The plot on the right side is
a typical outcome, exploiting
the $\pi\pi$ results for $X(2)$ and $Z(2)$ to determine the posterior
probability of the quark mass ratio $r$. We see that the current $\pi\pi$
data rule out small values of $r$ ($<15$) and favours slightly $r\sim 20$.
Including observables from the $N_f=3$ sector
($\pi K$ scattering, $\eta\to 3\pi$) within this framework
should finally elucidate the $N_f=2$ and $N_f=3$ chiral structures
of QCD vacuum.


\clearpage\label{tarasov}
\begin{center}
  {\Large{\bf On the role of multi-photon exchanges in the incoherent
      interaction of $\pi^+\pi^-$-atom with atoms of
      matter}}\footnote{The report was not presented because of a visa problem}

\bigskip

{\bf A.Tarasov and O.Voskresenskaya}\\[2mm]  

{\em Joint Institute for Nuclear Research, Dubna, Moscow Region,
  141980 Russia}
\end{center}


The Glauber theory for interaction of hydrogen-like elementary atoms
(EA) with target atoms (TA), developed in the papers [1-3], is
essentially based on the assumption that the Coulomb potential, created
by TA, does not change during the EA--TA interaction. In other words,
in this approximation all possible excitations of TA in intermediate
and/or in finale states are completely neglected. 

The Glauber theory corrected to account these effects is too
cumbersome. Here we list the simplest results of this theory
concerning the total cross sections of EA--TA interactions.

\begin{equation}
  \label{eq:1}
  \sigma^{tot}(i)=\sigma^{tot}_{coh}(i)+\sigma^{tot}_{incoh}(i)
\end{equation}

\begin{equation}
  \label{eq:2}
  \sigma^{tot}_{coh,incoh}(i)=\int d^3r |\Psi_i(\vec{r})|^2 d^2b 
  \Gamma(\vec{b},\vec{s})_{coh,incoh}
\end{equation}

\begin{equation}
  \label{eq:3}
  \Gamma(\vec{b},\vec{s})_{coh}=1-2\cos{[\Delta\chi(b,s)]}\exp{[-\Phi(b,s)/2]} 
  + \exp{[-\Phi(b,s)]}
\end{equation}

\begin{equation}
  \label{eq:4}
  \Gamma(\vec{b},\vec{s})_{incoh}=1 - \exp{[-\Phi(b,s)]}
\end{equation}

\begin{equation}
  \label{eq:5}
\Delta\chi(b,s)=\frac{2Z\alpha}{\beta} \int \frac{d^2q}{q^2} 
\left( e^{i\vec{q}\vec{b}_+} -e^{i\vec{q}\vec{b}_-}\right)
[1-S_1(\vec{q})]
\end{equation}

\begin{equation}
  \label{eq:6}
\Phi(b,s)=\frac{4Z\alpha^2}{\beta^2} \int \frac{d^2q_1}{q_1^2} \frac{d^2q_2}{q_2^2} 
\left( e^{i\vec{q_1}\vec{b}_+} -e^{i\vec{q}_1\vec{b}_-}\right)
\left( e^{-i\vec{q_2}\vec{b}_+} -e^{-i\vec{q}_2\vec{b}_-}\right)
W(\vec{q}_1,\vec{q}_2)
\end{equation}

\begin{equation}
  \label{eq:7}
W(\vec{q}_1,\vec{q}_2)=S_1(\vec{q}_1-\vec{q}_2)-S_1(\vec{q}_1)S_1(\vec{q}_2)+
(Z-1)[S_2(\vec{q}_1,\vec{q}_2)-S_1(\vec{q}_1)S_1(\vec{q}_2)]
\end{equation}

\begin{equation}
  \label{eq:8}
S_1(\vec{q})=\int d^3r e^{i\vec{q}\vec{r}} \rho_1(\vec{r})\,, \qquad
\int d^3r \rho_1(\vec{r})=1 
\end{equation}

\begin{equation}
  \label{eq:9}
S_2(\vec{q}_1,\vec{q}_2)=\int d^3r_1 d^3r_2 e^{i\vec{q}_1\vec{r}_1-i\vec{q}_2\vec{r}_2}
\rho(\vec{q}_1,\vec{q}_2)\,,
\qquad \int d^3r_2 \rho_2(\vec{r}_1,\vec{r}_2)=\rho_1(\vec{r}_1) 
\end{equation}

\begin{equation}
  \label{eq:10}
\vec{b}_{\pm} =\vec{b} \pm \vec{s}/2\,,\qquad  \vec{s}=\vec{r}_{\bot}
\end{equation}

\begin{equation}
  \label{eq:11}
W(\vec{q},\vec{q})=S_{incoh}(\vec{q})
\end{equation}
Here $\rho_{1,2}$ are one-particle and two-particle electron densities
of the target atom, $\sigma^{tot}_{coh,incoh}$ are the total cross
sections of EA--TA interactions without or with excitation of the
target atom. $\Phi$ in the above equations accounts the target atom
evitation both in intermediate and in final states. If one put
$\Phi=0$ then Eqs.(1-4) turns to the corresponding relations of papers
[1-3]. In particular in this limit $\sigma_{incoh}=0$. Relative
corrections to $\sigma^{tot}_{coh}$ caused by including intermediate
incoherent effects are of order
\begin{equation}
  \label{eq:12}
Z^3\alpha^4 \frac{\langle r^2\rangle_{EA}}{\langle r^2\rangle_{TA}} 
\ln{\left(\frac{\langle r^2\rangle_{EA}}{\langle r^2\rangle_{TA}}\right)} \ll 1
\end{equation}
and can be successfully neglected. The same is also true for all
partial coherent cross sections. From these estimations it follows
that the theory of papers [1-3] provides quite accurate description of
the coherent sector of EA--TA interactions. 

As for incoherent interactions, it follows from Eq.(4), that they can
be described by the Born approximation with the relative accuracy of
order 
\begin{equation}
  \label{eq:13}
Z\alpha^2 \frac{\langle r^2\rangle_{EA}}{\langle r^2\rangle_{TA}} 
\ln{\left(\frac{\langle r^2\rangle_{EA}}{\langle r^2\rangle_{TA}}\right)} \,.
\end{equation}

The results of performed analysis can be summarized as follows:
\begin{enumerate}
\item for description of the coherent interactions of EA with TA it is
  enough to use the simplified version of the Glauber theory
  neglecting effect of intermediate excitation of TA;
\item for description of the in coherent interactions of EA with TA it is
  enough to use the Born approximation. Just such prescriptions based
  on an intuitive consideration have been proposed by authors of paper [3].
\end{enumerate}


\clearpage\label{leutwyler}\begin{center}
{\Large{\bf Concluding remarks}}

\bigskip

{\bf H.~Leutwyler}\\[2mm]  

{\em Institute for Theoretical Physics, University of Bern, Sidlerstr.~5,
  CH-3012 Bern, 
  Switzerland} 

\end{center}

In the first part of the talk, I illustrated the relevance of low energy pion
physics for our current understanding of the basic interactions, using the
example of the Standard Model prediction for the muon magnetic
moment. I then discussed the remarkable recent progress achieved in $\pi\pi$
scattering, which now has become a precision science: in combination with the
low energy theorems of Chiral Perturbation 
Theory for the S-wave scattering lengths, the dispersion relations for the
partial waves (Roy equations) pin down the 
scattering amplitude to an amazing degree of accuracy [1-3] -- for once in
strong interaction physics, theory is ahead of experiment. The recent
$K_{e_4}$ data from Brookhaven [4] offer a rough test of one of the hypotheses
that underly the theoretical framework, as they show that the quark condensate
indeed represents the leading 
order parameter of the spontaneously broken chiral symmetry. 
Generalized ChPT has served its purpose and can now be dismissed. 

The Roy equations involve two subtraction constants, which may
be identified with the two S-wave scattering lengths. $K_{e_4}$ decays are
sensitive to one particular combination of these. The forthcoming
results from NA48/2 are eagerly awaited as they will narrow down the
uncertainty in the experimental information about this combination [5].  
The low energy theorem for the scalar radius involves a different combination,
but this theorem has not been subject to experimental test. The importance of
such a test was thoroughly discussed at this meeting [6-8]. A somewhat more
precise determination of the I = 2 $\pi\pi$ phase shift may become possible
at the COMPASS detector at CERN [9]
and, as pointed out by L.~Montanet, there are
beautiful data on other processes with two-pion configurations in the final
state, such as $D\rightarrow 3\pi$, which may also help to narrow
down the experimental uncertainties.  

The ideal laboratory for exploring the low energy
properties of the pions, however, is the atom consisting of a pair
of charged pions, also referred to as pionium. The DIRAC
collaboration at CERN has demonstrated that it is possible to generate such
atoms and to measure their lifetime [10].
Since the physics of the bound state is well understood [11-13],
the low energy properties of the scattering amplitude can unambiguously be
determined by studying these. It would be most deplorable if this
beautiful project were aborted before its physics potential [14] is
tapped. In particular, 
pionium level splittings would offer a clean and direct measurement of the
second subtraction constant. Data on $\pi K$
atoms would also be very valuable, as they would allow us to explore the role
played by the strange quarks in the QCD vacuum 
[15,16].  

Experimentally, pionic hydrogen has been explored very successfully [17]. Also,
there is significant progress in the physics of the bound state [18-20]. The
main problem here is that the ChPT predictions for the
scattering lengths have large uncertainties. At this time, accurate
results can only be obtained in the unphysical region of $\pi N$
scattering. Dispersion theory is needed to
confront these with data at or above threshold. An analogue of the Roy
equations has been formulated [21], but the framework yet needs to be
implemented. A comprehensive dispersive analysis of the $\pi N$ scattering
amplitude was carried out by G.~H\"ohler and collaborators 20 years ago [22]. 
An update of that analysis is under way, but will take some time for completion
[23]. Currently, the Standard Model predictions for pionic hydrogen are
too uncertain to draw physics conclusions from the available,
beautiful data, concerning the size of the $\sigma$-term, for instance. The
same applies to the predictions for $K N$ atoms, which are under
investigation in the DEAR experiment at Frascati [24]. 

There is plenty of work to be done in the field of hadronic atoms. Quite a few
experimentally accessible quantities can be calculated from first 
principles. By confronting theory with experiment, we can
critically examine our understanding of the laws of nature. 


\end{document}